\documentclass{LMCS}

\usepackage{amssymb,amsmath,latexsym,enumerate,hyperref,proof}

\newcommand{\cal}{\mathcal}

\newcommand{\com}{\newcommand}

\newcommand{\irule}[2]
{$\frac{\mbox{\raisebox{.4ex}{#1}}}{\mbox{\raisebox{-.8ex}[1.5ex][1.5ex]{#2}}}$}

\newcommand{\fl}{\noindent}
\newcommand{\hair}{\hspace{2mm}}
\newcommand{\vair}{\relax}
\newcommand{\bq}{\begin{quote}}
\newcommand{\eq}{\end{quote}}
\newcommand{\bc}{\begin{center}}
\newcommand{\ec}{\end{center}}
\newcommand{\bi}{\begin{itemize}}
\newcommand{\ei}{\end{itemize}}
\newcommand{\be}{\begin{enumerate}}
\newcommand{\ee}{\end{enumerate}}
\newcommand{\ba}{\begin{array}}
\newcommand{\ea}{\end{array}}
\newcommand{\bd}{\begin{description}}
\newcommand{\ed}{\end{description}}
\newcommand{\bt}{\begin{tabular}}
\newcommand{\et}{\end{tabular}}
\newcommand{\bp}{\begin{program}\small}
\newcommand{\ep}{\end{program}}

\newcommand{\bdfn}{\begin{defi}}
\newcommand{\edfn}{\end{defi}}

\newcommand{\blem}{\begin{lem}}
\newcommand{\elem}{\end{lem}}

\newcommand{\bcor}{\begin{cor}}
\newcommand{\ecor}{\end{cor}}

\newcommand{\bprf}{\proof}

\newcommand{\eprf}{\qed}

\newcommand{\bthm}{\begin{thm}}
\newcommand{\ethm}{\end{thm}}

\newcommand{\ignoreIF}[1] {}

\newcommand{\ignorePROD}[1] {}

\newcommand{\ignoreBOOL}[1] {}

\renewcommand{\int} {{\bf int}}

\com{\deq}{{\downarrow}{\raisebox{1ex}{\!\rmfamily\hspace{-0.8ex}\scriptsize=}}}
\com{\deqsm}{\downarrow{}^{\hspace{-1.1ex}=}} 

\newcommand{\programenvironment}{\programmode%
	\def\par{\leavevmode\endgraf}\obeylines\nobreak%
	\programmode}

\newcommand{\programmode}{\tt
	\catcode`\_=12 \catcode`\?=12 \catcode`\.=12 \catcode`\,=12
	\catcode`\;=12 \catcode`\:=12 \catcode`\@=12 \catcode`\~=12
        \catcode`\#=12 \catcode`\&=12      
	\obeyspaces\frenchspacing}%

\newenvironment{programintext}{\programenvironment}{}

\newenvironment{program}{\setlength{\partopsep}{0mm}\setlength{\topsep}{0mm}
	\begin{trivlist}\item[]
	\hspace*{5mm}\begin{minipage}{1.0\textwidth}
	\vspace{1mm}
	\begin{programintext}
	}{\end{programintext}
	\vspace{1mm}
	\end{minipage}
	\end{trivlist}
	\noindent}

{\catcode`\^^I=\active \obeyspaces\global\def^^I{ }}
{\obeyspaces\global\let =\ } 
{\catcode`\`=\active \gdef`{\relax\lq}}

\def\doframeit#1{\vbox{%
  \hrule height\fboxrule
    \hbox{%
      \vrule width\fboxrule \kern\fboxsep
      \vbox{\kern\fboxsep #1\kern\fboxsep }%
      \kern\fboxsep \vrule width\fboxrule }%
    \hrule height\fboxrule }}

\def\frameit{\smallskip \advance \linewidth by -7.5pt \setbox0=\vbox \bgroup
\strut \ignorespaces }

\def\endframeit{\ifhmode \par \nointerlineskip \fi \egroup
\doframeit{\box0}}

\theoremstyle{definition}

\newcounter{figur}
\newsavebox{\fighack}

\newenvironment{fig}[3]
{
\refstepcounter{figur}
\label{#3}
\sbox{\fighack}{{\it Figure \ref{#3}: #1}}
\begin{figure*}[#2]
\begin{frameit}
\hspace*{3mm}\begin{minipage}{0.95\textwidth}}{\end{minipage}            
\end{frameit}                                    
\usebox{\fighack} 
\end{figure*}}

\newenvironment{fig0}[3]
{
\refstepcounter{figur}
\label{#3}
\sbox{\fighack}{{\it Figure \ref{#3}: #1}}
\begin{figure}[#2]
\hspace*{0.02\textwidth}\begin{minipage}{0.98\textwidth}}{\end{minipage} 
\usebox{\fighack}\end{figure}}

\def\verbatim{\small \@verbatim \frenchspacing\@vobeyspaces \@xverbatim}

\newcommand{\tosub}[1]{\raisebox{-6.0pt}%
    {$\ \stackrel{\textstyle\to}{\vrule height4.5 pt width0 pt
    depth2.5 pt \smash{\scriptstyle#1}}\ $}}

\def\doi{4 (1:3) 2008}
\lmcsheading%
{\doi}
{1--39}
{}
{}
{Jul.~\phantom{0}5, 2007}
{Mar.~14, 2008}
{}   

\begin{document}

\title{Call-by-value Termination in the Untyped $\lambda$-calculus}

\author[N.~D.~Jones]{Neil D.~Jones\rsuper a}
\address{{\lsuper a}DIKU, University of Copenhagen, Denmark}
\email{neil@diku.dk}

\author[N.~Bohr]{Nina Bohr\rsuper b}
\address{{\lsuper b}IT-University of Copenhagen, Denmark}
\email{nina@itu.dk}

\keywords{Program analysis, Termination analysis, Untyped Lambda
  calculus, The Size-Change Principle} \subjclass{F.3.2, D.3.1}


\begin{abstract}
A fully-automated algorithm is developed able to show that evaluation
of a given untyped $\lambda$-expression will terminate under CBV
(call-by-value).  The ``size-change principle'' from first-order
programs is extended to arbitrary untyped $\lambda$-expressions in two
steps.  The first step suffices to show CBV termination of a single,
stand-alone $\lambda$-expression. The second suffices to show CBV
termination of any member of a regular set of $\lambda$-expressions,
defined by a tree grammar.  (A simple example is a minimum function,
when applied to arbitrary Church numerals.)  The algorithm is sound
and proven so in this paper. The Halting Problem's undecidability
implies that any sound algorithm is necessarily incomplete: some
$\lambda$-expressions may in fact terminate under CBV evaluation, but
not be recognised as terminating.

The intensional power of the termination algorithm is reasonably
high. It certifies as terminating many interesting and useful general
recursive algorithms including programs with mutual recursion and
parameter exchanges, and Colson's ``minimum'' algorithm.  Further, our
type-free approach allows use of the Y combinator, and so can identify
as terminating a substantial subset of PCF.

\end{abstract}

\maketitle
\bigskip
\bigskip

\tableofcontents\newpage

\section{Introduction}

\noindent The {\em size-change} analysis by Lee, Jones and Ben-Amram
\cite{popl01} can show termination of programs whose parameter values
have a well-founded size order.  The method is reasonably general,
easily automated, and does not require human invention of lexical or
other parameter orders.  It applies to first-order functional
programs. This paper applies similar ideas to termination of {\em
higher-order} programs. For simplicity and generality we focus on the
simplest such language, the $\lambda$-calculus.

\subsection*{Contribution of this paper}

Article \cite{rta} (prepared for an invited conference lecture) showed
how to lift the methods of \cite{popl01} to show termination of closed
$\lambda$-expressions.  The current paper is a journal version of
\cite{rta}. It extends \cite{rta} to deal not only with a single
$\lambda$-expression in isolation, but with a regular set of
$\lambda$-expressions generated by a finite tree grammar.  For
example, we can show that a $\lambda$-expression terminates when
applied to Church numerals, even though it may fail to terminate on
all possible arguments.  This paper includes a number of examples
showing its analytical power, including programs with primitive
recursion, mutual recursion and parameter exchanges, and Colson's
``minimum'' algorithm.  Further, examples show that our type-free
approach allows free use of the Y combinator, and so can identify as
terminating a substantial subset of PCF.

\subsection{Related work}

Jones \cite{Jones:FALambda} was an early paper on control-flow analysis of the untyped $\lambda$-calculus.
Shivers' thesis and subsequent work  \cite{shivers1991,shivers2004} on CFA (control flow analysis) 
developed this approach considerably further and applied it to the Scheme
programming language. This line is
closely related to the approximate semantics (static control graph) 
of Section \ref{sec-abstract-interpretation-for-cfg}
 \cite{Jones:FALambda}.

\subsubsection*{Termination of untyped programs} 

Papers based on \cite{popl01} have used size-change graphs to 
find bounds on program running times 
(Frederiksen and Jones \cite{frederiksen-jones}); solved related problems, e.g., to ensure that 
partial evaluation will
terminate  (Glenstrup and Jones, Lee \cite{glenstrup-jones,lee-ptime-bta}); and
found more efficient (though less precise)
algorithms (Lee \cite{lee-poly-time-analysis}).
Further, Lee's thesis  \cite{lee-thesis} extends the first-order size-change method  
\cite{popl01} to handle higher-order named combinator programs.
It uses a different approach than ours, and appears to be less general.

We had anticipated from the start that our framework could
naturally be extended to higher-order functional programs, 
e.g., functional subsets of Scheme or ML. This has since been 
confirmed by Sereni and Jones, first reported in  \cite{serenijones}.
Sereni's Ph.D.\ thesis  \cite{sereni} develops this direction in considerably more detail
with full proofs, and also investigates problems with lazy (call-by-name) languages. 
Independently and a bit later, Giesl and coauthors have addressed the analysis of 
the lazy functional language Haskell \cite{haskell}.

\subsubsection*{Termination of typed $\lambda$-calculi} 

Quite a few people have written  about termination based on types. 
Various subsets of the
$\lambda$-calculus, in particular subsets typable by various 
disciplines,  have been proven
strongly normalising. 
Work in this direction includes 
pathbreaking results by Tait  \cite{tait} and others
concerning simple types,
and
Girard's System F  \cite{girard}.
Abel, Barthe and others have done newer type-based approaches to show termination
of a $\lambda$-calculus extended with recursive data types \cite{abel,abelthesis,barthe}.

{\em Typed functional languages:}
Xi's Ph.D.\ research focused on tracing value flow via data types
for termination verification 
in higher order programming languages  \cite{xi}, 
Wahlstedt has an approach to combine size-change termination 
analysis with constructive type theory
\cite{wahlstedt, wahlstedt1}.

{\em Term rewriting systems:}
The popular ``dependency pair'' method was developed
 by Arts  and Giesl \cite{giesl-arts} for first-order programs in TRS form.
This community has begun to study termination  of
higher order 
term rewriting systems, including research by 
Giesl et.al.\ \cite{giesl-thiemann,haskell},
Toyama \cite{toyama} and others.

\section{The call-by-value $\lambda$-calculus}

First, we review relevant definitions and results for the
call-by-value $\lambda$-calculus, and then provide an observable
characterisation of the behavior of a nonterminating expression.

\subsection{Classical semantics}
\label{classical-semantics}

\bdfn
\label{def-syntax}
  {\it Exp} is the set of all $\lambda$-expressions that can be 
formed by these syntax rules, where {\tt @} 
is the {\em application operator} (sometimes omitted). We 
 use the {\tt teletype} font for $\lambda$-expressions.
\vair

{\tt 
\bt{lcl}
e, P &\ ::=\ &\  x | e @ e | $\lambda$x.e\\
x &\ ::=\ &\  {\rm Variable name}
\et}
\be[$\bullet$]
\item 
The set of {\em free variables} ${\it fv}({\tt e})$ is defined as usual: ${\it fv}({\tt x})=\{{\tt x}\}$, 
${\it fv}({\tt e @ e}')={\it fv}({\tt e})\cup{\it fv}({\tt e}')$ 
and 
${\it fv}(\lambda{\tt x.e})={\it fv}({\tt e})\setminus\{{\tt x}\}$.
 A {\it closed}  $\lambda$-expression {\tt e} satisfies 
${\it fv}({\tt e}) = \emptyset$.

\item
A {\em program}, usually 
denoted by {\tt P},  is any closed $\lambda$-expression. 

\item The set of {\em subexpressions} of a 
$\lambda$-expression {\tt e} is denoted by ${\it subexp}({\tt e})$.
\ee
\edfn

The following is standard, e.g., \cite{plotkin}. Notation: 
$\beta$-reduction is done by substituting 
$v$ for all free occurrences of {\tt x} in {\tt e},
written
${\tt e}[v/{\tt x}]$,  and renaming 
$\lambda$-bound variables if needed to avoid 
capture.

\bdfn \label{def-call-by-value-evaluation}
{\rm (Call-by-value semantics)}
The {\em call-by-value evaluation relation}  is defined by the 
following inference rules, with judgement form ${\tt e}\Downarrow v$
where ${\tt e}$ is a closed $\lambda$-expression and $v \in {\it ValueS}$.
{\it ValueS} (for ``standard value'') is the set of all abstractions 
$\lambda{\tt x.e}$. \medskip

$$
\infer[\mbox{{\it {\small If v $\in$ ValueS}}\ }(\textnormal{ValueS})]{v\Downarrow v}{}
\qquad
\infer[(\textnormal{ApplyS})]{{\tt e}_1{\tt@e}_2\Downarrow v}{{\tt e}_1\Downarrow \lambda{\tt x.e}_0 
      & 
      {\tt e}_2\Downarrow v_2 
      &
{\tt e}_0[v_2/{\tt x}]\Downarrow v
}
$$
\edfn
\bigskip

\blem\label{lem-determinism}
{\rm(Determinism)} If ${\tt e}\Downarrow v$ and ${\tt e}\Downarrow 
w$  then $v=w$.
\elem


\subsection{Nontermination is sequential} 
\label{sec-making-nontermination-visible}

A proof of ${\tt e}\Downarrow v$ is a finite object, and no such proof
exists if the evaluation of ${\tt e}$ fails to terminate.  Thus in
order to be able to trace an arbitrary computation, terminating or
not, we introduce a new ``calls'' relation ${\tt e}\to {\tt e}'$, in
order to make nontermination visible.\nobreak

\subsubsection*{The ``calls'' relation}

The rationale is straightforward: ${\tt e}\to {\tt e}'$ if in order to
deduce ${\tt e}\Downarrow v$ for some value $v$, it is necessary first
to deduce ${\tt e}'\Downarrow u$ for some $u$, i.e., some inference
rule has form \irule{$\ldots\ {\tt e}'\Downarrow \ ?\ \ldots$}{${\tt
e}\Downarrow\ ?$}.  Applying this to Definition
\ref{def-call-by-value-evaluation} gives the following.\break

\bdfn \label{def-evaluation-and-call} {\rm (Evaluation and call semantics)} 
The {\em evaluation 
and call relations}   
are defined by the following inference rules, where
$\tosub{r}, \tosub{d}, \tosub{c}\ 
\subseteq{\it Exp}\times{\it Exp}$\footnote{Naming: $r,d$ in 
$\tosub{r},\tosub{d}$ are the last letters of {\em operato{\bf r}} 
and {\em 
operan{\bf d}}, and $c$ in $\tosub{c}$ stands for ``call''. }.
\medskip

$$
\infer[\mbox{{\it {\small If v $\in$ ValueS}}\ }(\textnormal{Value})]{v\Downarrow v}{}
\qquad
\infer[(\textnormal{Operator})]{{\tt e}_1{\tt@e}_2\tosub{r}{\tt e}_1}{}
\qquad
\infer[(\textnormal{Operand})]{{\tt e}_1{\tt@e}_2\tosub{d}{\tt e}_2}{{\tt e}_1\Downarrow v_1
}
$$

$$
\infer[(\textnormal{Call$_0$})]{{\tt e}_1{\tt@e}_2\tosub{c}{\tt e}_0[v_2/{\tt x} ]}{{\tt e}_1\Downarrow \lambda{\tt x.e}_0
 & {\tt e}_2\Downarrow v_2}
\qquad
\infer[(\textnormal{Apply$_0$})]{{\tt e}_1{\tt@e}_2\Downarrow v}{{\tt e}_1\Downarrow \lambda{\tt x.e}_0
& 
 {\tt e}_2\Downarrow v_2
&
{\tt e}_0[v_2/{\tt x} ]\Downarrow v}
$$

\edfn
\bigskip

\fl 
For convenience we will sometimes 
combine the three into a single {\em call relation} 
$\to\ =\tosub{r}\cup\tosub{d}\cup\tosub{c}$.
As usual, we write $\to^+$ for the transitive closure of $\to$, and 
 $\to^*$ for its reflexive transitive closure. 
We will sometimes write $s\Downarrow$ to mean $s\Downarrow 
v$ for some $v\in{\it ValueS}$, and  write $s\not\Downarrow$ to mean 
there is no $v\in{\it ValueS}$ such that $s\Downarrow 
v$, i.e., if evaluation of $s$ does not terminate.

\subsubsection*{A small improvement to the operational semantics} 

Note
that rules (Call$_0$)
and (Apply$_0$) from Definition \ref{def-evaluation-and-call} overlap:
${\tt e}_2\Downarrow v_2$ appears in both, as does
${\tt e}_0[v_2/{\tt x} ]$. Thus (Call$_0$) can be used 
as an intermediate step 
to simplify (Apply$_0$), giving a more orthogonal set of  rules. 
Variations on the following  
combined set will be used in the rest of the paper:

\bdfn \label{def-evaluate-and-call-relation}
(Combined evaluate and call rules, standard semantics)
\medskip

$$
\infer[\mbox{{\it {\small If v $\in$ ValueS}}\ }(\textnormal{Value})]{v\Downarrow v}{}
\qquad
\infer[(\textnormal{Operator})]{{\tt e}_1{\tt@e}_2\tosub{r}{\tt e}_1}{}
\qquad
\infer[(\textnormal{Operand})]{{\tt e}_1{\tt@e}_2\tosub{d}{\tt e}_2}{{\tt e}_1\Downarrow v_1
}
$$

$$
\infer[(\textnormal{Call})]{{\tt e}_1{\tt@e}_2\tosub{c}{\tt e}_0[v_2/{\tt x} ]}{{\tt e}_1\Downarrow \lambda{\tt x.e}_0
 & {\tt e}_2\Downarrow v_2}
\qquad
\infer[(\textnormal{Apply})]{{\tt e}_1{\tt@e}_2\Downarrow v}
{{\tt e}_1{\tt@e}_2\tosub{c}{\tt e}'
&
{\tt e}'\Downarrow v}
$$
\edfn
\smallskip

\fl The {\em call tree} of program {\tt P} is the smallest set of
 expressions $CT$ containing {\tt P} that is closed under $\to$ .  It
 is not necessarily finite.

\blem \label{lem-NIS-standard} {\rm (}NIS, or {\bf\underline N}ontermination 
{\bf\underline I}s 
{\bf\underline S}equential{\rm )} 
Let {\tt P} be a program. 
Then ${\tt P}\Downarrow$  if and only  if $CT$ has no infinite call chain starting with ${\tt P}$:
$$
{\tt P} = {\tt e}_0\to{\tt e}_1\to{\tt e}_2\to \ldots 
$$
\elem

\fl{\em Example:} evaluation of expression
$\Omega=(\lambda{\tt x.x@x}){\tt@}(\lambda{\tt y.y@y})$ 
yields an infinite call chain:
$$
\Omega=(\lambda{\tt x.x@x}){\tt@}(\lambda{\tt y.y@y})\to(\lambda{\tt y.y@y}){\tt@}(\lambda{\tt y.y@y})\to
  (\lambda{\tt y.y@y}){\tt@}(\lambda{\tt y.y@y})\to\ldots
$$
By the NIS Lemma all nonterminating computations
give rise to   {\em infinite linear call chains}.
Such call chains need not, however, be repetitive as in this example, 
or even finite.

Informally ${\tt e}_0\not\Downarrow$ 
implies existence of an infinite call chain as follows:
Try to build,  
bottom-up and left-to-right,  a 
proof tree for ${\tt e}_0\Downarrow v$.
Since call-by-value evaluation cannot ``get stuck''
this process will continue infinitely, leading to an infinite call chain.
Figure \ref{fig-NIS-behavior} shows 
such a call tree with infinite path starting with ${\tt e}_0\to{\tt e}_1\to{\tt e}_2\to{\tt e}_3\to\ldots$, 
where $\to\ =\tosub{r}\cup\tosub{d}\cup\tosub{c}$.
The Appendix contains a formal proof.

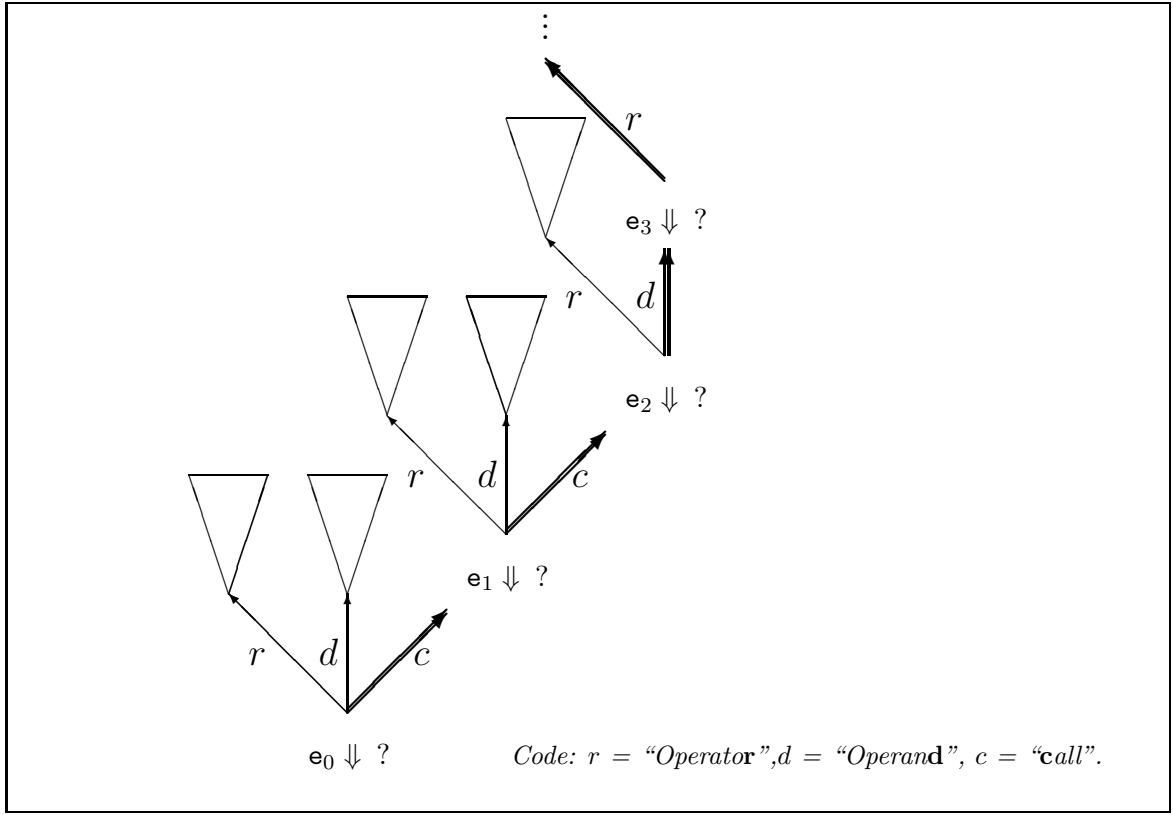
\begin{fig}{Nontermination implies existence of an infinite call chain}{bth}{fig-NIS-behavior}

\fl{\tt    \setlength{\unitlength}{1.5pt}
\begin{picture}(530,200)(0,-10)
\thinlines
	      
              \put(150,148){\vector(-1,1){30}}

              \put(150,103){\vector(-1,1){30}}
              \put(130,163){\line(-1,0){20}}
              \put(120,133){\line(-1,3){10}}
              \put(120,133){\line(1,3){10}}
	      
              \put(110,58){\vector(-1,1){30}}
              \put(90,118){\line(-1,0){20}}
              \put(80,88){\line(-1,3){10}}
              \put(80,88){\line(1,3){10}}
	      
              \put(110,58){\vector(0,1){30}}
              \put(120,118){\line(-1,0){20}}
              \put(110,88){\line(-1,3){10}}
              \put(110,88){\line(1,3){10}}
	      
              \put(70,13){\vector(-1,1){30}}
              \put(50,73){\line(-1,0){20}}
              \put(110,88){\line(-1,3){10}}
              \put(40,43){\line(1,3){10}}
	      
              \put(70,13){\vector(-1,1){30}}
              \put(50,73){\line(-1,0){20}}
              \put(40,43){\line(-1,3){10}}
              \put(40,43){\line(1,3){10}}
	      
              \put(70,13){\vector(0,1){30}}
              \put(80,73){\line(-1,0){20}}
              \put(70,43){\line(-1,3){10}}
              \put(70,43){\line(1,3){10}}

\thicklines
	      
              \put(140,135){${\tt e}_3 \Downarrow\  ?$
}
              \put(140,160){\Large $r$}
	      
              \put(140,90){${\tt e}_2 \Downarrow\  ?$}
              \put(150,103){\vector(0,1){27}}
              \put(151,103){\vector(0,1){27}}
              \put(125,115){\Large $r$\hspace{7mm}$d$}
	      
              \put(100,45){${\tt e}_1 \Downarrow\  ?$}
              \put(110,58){\vector(1,1){25}}
              \put(110,59){\vector(1,1){25}}
              \put(85,70){\Large $r$\hspace{7mm}$d$\hspace{10mm}$c$}
	      
              \put(60,0){${\tt e}_0 \Downarrow\  ?$
\hspace{14mm}{\it{\small Code: $r$ = ``Operato{\bf r}'',$d$ = ``Operan{\bf d}'',  
 $c$ = ``{\bf c}all''.}}}
              \put(70,13){\vector(1,1){25}}
              \put(70,14){\vector(1,1){25}}
              \put(45,25){\Large $r$\hspace{7mm}$d$\hspace{10mm}$c$}
	      
              \put(150,147){\vector(-1,1){30}}
              \put(150,148){\vector(-1,1){30}}
              \put(118,183){$\vdots$}

\end{picture}}

\end{fig}

\section{An approach to termination analysis}

 The ``size-change termination'' analysis of  Lee, Jones and Ben-Amram \cite{popl01} is based on 
several concepts,  including:
\be[(1)]
\item\label{chal-identifying-nontermination}
Identifying nontermination as caused by 
{\em infinitely long sequential state transitions}.

\item\label{chal-control-points} 
A fixed set of {\em program control points}.

\item\label{chal-observable-decreases}
{\em Observable decreases} in data value sizes.

\item\label{chal-construct-size-change-graph}
{\em  Construction} of  one size-change graph for each 
 function call.

\item\label{chal-control-flow-graph}
Finding the program's  {\em control flow graph}, and the 
call sequences that follow it.

\ee

\fl The NIS Lemma establishes point \ref{chal-identifying-nontermination}. However, concepts 
\ref{chal-control-points},
\ref{chal-observable-decreases},
\ref{chal-construct-size-change-graph} and
\ref{chal-control-flow-graph}
all seem a priori absent from the 
$\lambda$-calculus, 
except that an application must be a call; and even then, 
it is not a priori clear {\em which} function is being called. 
We will show, one step at a time, that all the concepts do in fact 
exist in call-by-value  $\lambda$-calculus evaluation.

\subsection{An environment-based semantics}
\label{sec-environment-based-semantics}

Program flow analysis usually requires evident 
program control points.
An alternate environment-based formulation remedies their absence in the $\lambda$-calculus. 
The ideas were formalised  
by Plotkin \cite{plotkin},
and have  long been used in implementations 
of functional programming language such as
{\sc scheme} and {\sc ml}.

\bdfn \label{def-environment-etc} {\rm (States, etc.)} 
Define {\it State}, {\it Value}, {\it Env} to be the smallest 
sets such that
\vspace{1mm}

\bt{lccclr}

{\it State} &$\hair=\hair\{$&  ${\tt e}:\rho$ 
&$\hair|\hair$& ${\tt e}\in{\it Exp},
\rho\in{\it Env}\mbox{\ and \ }
{\it dom}(\rho) \supseteq {\it fv}({\tt e})$&$\}$\\

{\it Value} &$\hair=\hair\{$&  $\lambda{\tt x.e}:\rho$ 
&$\hair|\hair$& $
\lambda{\tt x.e}:\rho\in{\it State}$&$\}$\\

{\it Env} &$\hair=\hair\{$& $ \rho:X\to {\it Value}$ 
&$\hair|\hair$& $X\mbox{\  is a finite set of variables}$&$\}$
\et
\vair

Equality of states is defined by:
$${\tt e}_1:\rho_1 = {\tt e}_2:\rho_2\ \textnormal{ holds  if } \ {\tt e}_1 ={\tt e}_2  \textnormal{ and } \rho_1(x) = \rho_2(x)\textnormal{ for all }   {\tt x}\in fv({\tt e}_1)$$

\fl
The empty environment with  domain $X=\emptyset$ is written $[]$. 
The environment-based evaluation judgement form is 
$s\Downarrow v$ where $s\in{\it State}, v \in {\it Value}$. 
\edfn

\fl The Plotkin-style rules  follow the pattern of Definition 
\ref{classical-semantics}, except that 
substitution ($\beta$-reduction) 
${\tt e}_0[v_2/{\tt x}]$ of the (CallS) rule is replaced 
by a ``lazy 
substitution'' that just updates the environment in the new (Call) rule.
Further, variable values are fetched from the environment

\bdfn \label{def-environment-evaluation} {\rm (Environment-based
evaluation semantics)} The evaluation relation $\Downarrow,$ is
defined by the following inference rules.
\bigskip

$$
\infer[\mbox{{\it {\small If v $\in$ Value}}\ }(\textnormal{ValueE})]{v\Downarrow v}{}
\qquad
\infer[(\textnormal{VarE})]{{\tt x} :\rho\Downarrow\rho({\tt x})}{}
$$

$$
\infer[(\textnormal{ApplyE$_0$})]{{\tt e}_1{\tt@e}_2:\rho\Downarrow v}
{{\tt e}_1:\rho\Downarrow \lambda{\tt x.e}_0:\rho_0
 & 
   {\tt e}_2:\rho\Downarrow v_2
&
  {\tt e}_0:\rho_0[{\tt x}\mapsto v_2]\Downarrow v}
$$
\smallskip

\subsection{States are tree structures}

A state has form $s={\tt e:} \ \rho$ as in Definition 
\ref{def-environment-etc} where $\rho$ binds the free variables of 
{\tt e} to values, which are themselves states. 
Consider, for two examples,  these two states
\medskip

{\tt
\bt{lclcl}
$s$ & $=$ & {\tt e}:$\rho$ & $=$ & r@(r@a):$
  [{\tt  r}\mapsto{\tt succ:[]},
   {\tt  a}\mapsto{\tt \underline{2}}:[]]$ \\\\
$s'$ & $=$ & {\tt e}$'$:$\rho'$ & $=$ & r@(r@a):$
  [{\tt  r}\mapsto\lambda{\tt a\,.\,}{\tt r@(r@a)}:[{\tt  r}\mapsto{\tt 
  succ:[]}],
   {\tt  a}\mapsto{\tt \underline{2}}:[]]$ \\
\et
}
\medskip

\fl(written in our usual linear notation
and using the standard Church numerals 
${\tt \underline{0}},{\tt \underline{1}},{\tt \underline{2}},\ldots$. For brevity details of the successor 
function {\tt succ} are omitted.
It is straightforward to verify  that 
$s\Downarrow {\tt \underline{4}}$ and
$s'\Downarrow {\tt \underline{6}}$ by Definition \ref{def-environment-evaluation}.

More generally, each value bound in an environment is a state in turn, so in full 
detail a state's structure  is 
a {\em finite tree}. (The levels of this tree represent variable 
bindings, not to be confused with the syntactic or subexpression
tree structures from 
Figure \ref{fig-ex-call relation}.)

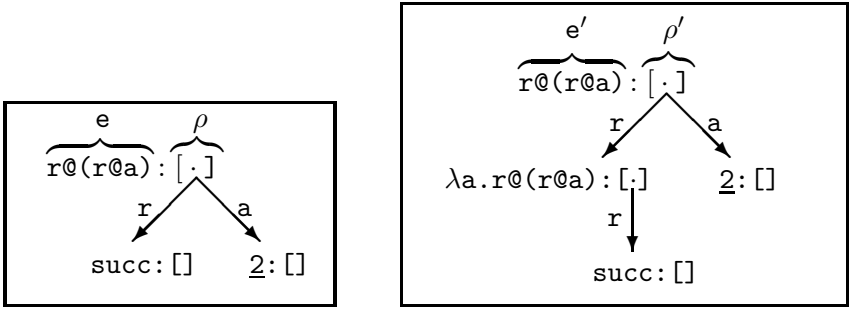
\begin{fig0}{Structures of two states $s,s'$. Each state is  a finite tree.}{htb}{fig-state-as-tree}

{\tt    \setlength{\unitlength}{0.8pt}
\begin{picture}(292,160)(-10,60)
\thicklines    
              \put(-15,70){\framebox(155,95){}}

              \put(27,154){${\tt e}\hspace{11mm}\rho$}
              \put(-3,131){
$\overbrace{\mbox{\tt r@(r@a)}}{\tt:}
 \overbrace{{\tt[}\cdot\mbox{\tt]}} $}
              \put(75,130){\vector(1,-1){30}}
              \put(75,130){\vector(-1,-1){30}}
              \put(47,112){{\tt r}}
              \put(94,112){{\tt a}}
              \put(25,85){{\tt succ:[]}}
              \put(100,85){{\tt \underline{2}:[]}}

\end{picture}}
{\tt    \setlength{\unitlength}{0.8pt}
\begin{picture}(292,160)(250,20)
\thicklines 
              \put(130,30){\framebox(210,142){}} 
	      
              \put(207,156){${\tt e}'\hspace{10mm}\rho'$}  

              \put(185,131){$\overbrace{\mbox{\tt r@(r@a)}}{\tt:}
 \overbrace{{\tt[}\cdot\mbox{\tt]}} $}
              \put(255,130){\vector(1,-1){30}}
              \put(255,130){\vector(-1,-1){30}}
              \put(228,112){{\tt r}}
              \put(274,112){{\tt a}}
              \put(150,85){{\tt $\lambda$a.r@(r@a):[$\cdot$]}}
              \put(280,85){{\tt {\underline{2}}:[]}}

              \put(239,85){\vector(0,-1){30}}
              \put(227,67){{\tt r}}

              \put(220,41){{\tt succ}:[]}

\end{picture}}

\end{fig0}

\noindent Figure \ref{fig-state-as-tree} shows the structure of these two states, with abbreviations for  Church numerals such as
${\tt \underline{2}} = \lambda {\tt s}\lambda {\tt z}\,.\,{\tt 
s@(s@z)}$.

\medskip

\edfn

\fl

\subsection{Nontermination made visible in an environment-based semantics}

Straightforwardly adapting the approach of 
Section \ref{sec-making-nontermination-visible}.
gives  the following  set of inference rules, variations on which will be used in the rest of the paper:

\bdfn \label{def-evaluate-and-call-relation-environment}
(Combined evaluate and call rules, environment semantics)
\bigskip

$$
\infer[\mbox{{\it {\small If v $\in$ Value}}\ }(\textnormal{Value})]{v\Downarrow v}{}
\qquad
\infer[(\textnormal{Var})]{{\tt x} :\rho\Downarrow\rho({\tt x})}{}
$$

$$
\infer[(\textnormal{Operator})]{{\tt e}_1{\tt@e}_2:\rho\tosub{r}{\tt e}_1:\rho}{}
\qquad
\infer[(\textnormal{Operand})]{{\tt e}_1{\tt@e}_2:\rho\tosub{d}{\tt e}_2:\rho}{{\tt e}_1:\rho\Downarrow v_1}
$$

$$
\infer[(\textnormal{Call})]{{\tt e}_1{\tt@e}_2:\rho\tosub{c}{\tt e}_0:\rho_0[{\tt x}\mapsto v_2]}
{{\tt e}_1:\rho\Downarrow \lambda{\tt x.e}_0:\rho_0
 & {\tt e}_2:\rho\Downarrow v_2}
\qquad
\infer[(\textnormal{Apply})]{{\tt e}_1{\tt@e}_2:\rho\Downarrow v}{{\tt e}_1{\tt@e}_2:\rho\tosub{c}{\tt e}':\rho' &
{\tt e}':\rho'\Downarrow v}
$$
\edfn
\bigskip

\noindent The following is proven in the same way as Lemma  \ref{lem-NIS-standard}.

\blem \label{lem-NIS-environment} {\rm (}NIS, or {\bf\underline N}ontermination 
{\bf\underline I}s 
{\bf\underline S}equential{\rm )} 
Let {\tt P} be a program. 
Then ${\tt P}:[]\Downarrow$  if and only if $CT$ has no infinite call chain staring with ${\tt P}:[]$
(where $\to\ =\tosub{r}\cup\tosub{d}\cup\tosub{c}$):
$$
{\tt P}:[] = {\tt e}_0:\rho_0\to{\tt e}_1:\rho_1\to{\tt e}_2:\rho_2\to \ldots 
$$
\elem
\medskip

Following the lines of Plotkin \cite{plotkin}, the environment-based semantics is 
shown equivalent to the usual semantics in 
the sense that they have the same termination behaviour. Further, 
when evaluation terminates the computed values are related by function
$F:States\to Exp$ defined by
$$F({\tt e}:\rho) = {\tt e}[F(\rho({\tt x}_1))/ {\tt x}_1,...,F(\rho({\tt x}_k))/ {\tt x}_k]
\mbox{\ \ where  }
\{{\tt x}_1,.., {\tt x}_k\} = 
fv({\tt e}).$$

\vair
\blem\label{lem-equivalent-semantics} 
${\tt P}:[] \Downarrow v$ (by Definition \ref{def-environment-evaluation}) 
implies
${\tt P}\Downarrow F(v)$  (by Definition \ref{def-evaluate-and-call-relation}), and\\
${\tt P}\Downarrow w$ implies there exists $v'$ such that  ${\tt P}:[] \Downarrow v'$ and $F(v')=w$.
\elem 

\medskip

\fl{\em Example:} evaluation of closed 
$\Omega=(\lambda{\tt x.x@x}){\tt@}(\lambda{\tt y.y@y})$ 
yields an infinite call chain:
$$
\Omega:[]=(\lambda{\tt x.x@x}){\tt@}(\lambda{\tt y.y@y}):[]\to
{\tt x@x}:\rho_1\to
  {\tt y@y}:\rho_2\to{\tt y@y}:\rho_2\to{\tt y@y}:\rho_2\to\ldots
$$
where $\rho_1=[{\tt x}\mapsto\lambda{\tt y.y@y}:[]]$\hair and \hair
	      $\rho_2=[{\tt y}\mapsto\lambda{\tt y.y@y}:[]]$.

\subsection{A control point is a subexpression of a $\lambda$-expression}

The following {\em subexpression property} 
does not hold for the classical rewriting $\lambda$-calculus 
semantics, but does hold for Plotkin-style environment semantics of 
Definition \ref{def-environment-evaluation}. It is central to our program 
analysis: A {\em control point} will be a
subexpression of the program {\tt P} being analysed, and our analyses will trace 
program information flow to and from subexpressions of {\tt P}.

\blem \label{lem-prog-subexpression}
If 
${\tt P}:[]\Downarrow \lambda{\tt x.e}:\rho$ then 
$\lambda{\tt x.e}\in{\it subexp}({\tt P})$. {\rm [Recall Definition \ref{def-syntax}.]}
\elem
This is proven as follows, using a more general inductive hypothesis.

\bdfn The {\em expression support} of a given state $s$ is
 ${\it exp\_sup}(s)$, defined by
\[{\it exp\_sup}({\tt e}:\rho) = {\it subexp}({\tt e}) \cup  
  \bigcup_{{\tt x}\in{\it fv}({\tt e})}
     {\it exp\_sup}(\rho({\tt x}))
     \]
\edfn

\blem \label{lem-subexpression} 
{\rm (Subexpression property)} If $s\Downarrow s'$ or $s\to s'$ then
${\it exp\_sup}(s)\supseteq {\it exp\_sup}(s')$. 
\elem

\bprf 
This follows by induction on the proof of 
 $s\Downarrow v$ or $s\to s'$. Lemma \ref{lem-prog-subexpression} is 
 an immediate corollary.

Base cases: $s={\tt x}:\rho$ and $s=\lambda{\tt x.e}:\rho$ 
 are immediate.
 For rule (Call)  suppose 
 ${\tt e}_1:\rho\Downarrow \lambda{\tt x.e}_0:\rho_0$ and
 ${\tt e}_2:\rho\Downarrow v_2$. By induction
$${\it exp\_sup}({\tt e}_1:\rho)\supseteq
  {\it exp\_sup}(\lambda{\tt x.e}_0:\rho_0)
  \mbox{\hair and\hair}{\it exp\_sup}({\tt e}_2:\rho)\supseteq
  {\it exp\_sup}(v_2)$$
Thus

$\ba{cclll}
{\it exp\_sup}({\tt e}_1{\tt @e}_2:\rho) & \supseteq &
  {\it exp\_sup}({\tt e}_1:\rho)\cup{\it exp\_sup}({\tt e}_2:\rho)
   & \supseteq\\

  {\it exp\_sup}(\lambda{\tt x.e}_0:\rho_0) \cup{\it 
  exp\_sup}(v_2) & \supseteq &

  {\it exp\_sup}({\tt e}_0:\rho_0[{\tt x}\mapsto v_2]) \\
  \ea
$
\medskip

\fl For rule (Apply) we have 
${\it exp\_sup}({\tt e}_1{\tt @e}_2:\rho)\supseteq
  {\it exp\_sup}({\tt e}':\rho') \supseteq
  {\it exp\_sup}(v)$.  The cases (Operator), (Operand) are immediate.
  \eprf

\subsection{Finitely describing a program's  computation space }
\label{sec-lambda-lacks}

A standard approach to program analysis is to trace data flow along the arcs of 
 the program's  {\em dynamic control graph} or DCG. In our case this is  the call relation $\to$ of
Definition \ref{def-evaluate-and-call-relation}.
Unfortunately the DCG may be infinite, so for program analysis we will 
instead compute a 
safe finite approximation called the SCG, for  {\em static control graph}.

\begin{exa}

Figure \ref{fig-ex-call relation} shows the combinator  
$\Omega=(\lambda{\tt x.x@x}){\tt@}(\lambda{\tt y.y@y})$ as a 
syntax tree whose  subexpressions are labeled by numbers. To its right is the ``calls'' 
relation $\to$. It has an infinite call chain:
$$
\Omega:[]\to{\tt x@x}:\rho_1\to
  {\tt y@y}:\rho_2\to{\tt y@y}:\rho_2\to{\tt y@y}:\rho_2\to\ldots
$$
Using subexpression numbers, the loop is
$$
{\tt 1}:[]\to{\tt 3}:\rho_1\to{\tt 7}:\rho_2\to{\tt 7}:\rho_2\to\ldots
$$
where $\rho_1=[{\tt x}\mapsto\lambda{\tt y.y@y}:[]]$\hair and \hair
	      $\rho_2=[{\tt y}\mapsto\lambda{\tt y.y@y}:[]]$.	      
The  set of states reachable from ${\tt P}:[]$ is 
finite, so this computation is in fact a ``repetitive loop.'' 
(It is also possible that
a computation will reach infinitely many  states that are all 
different.)
\end{exa}
\bigskip

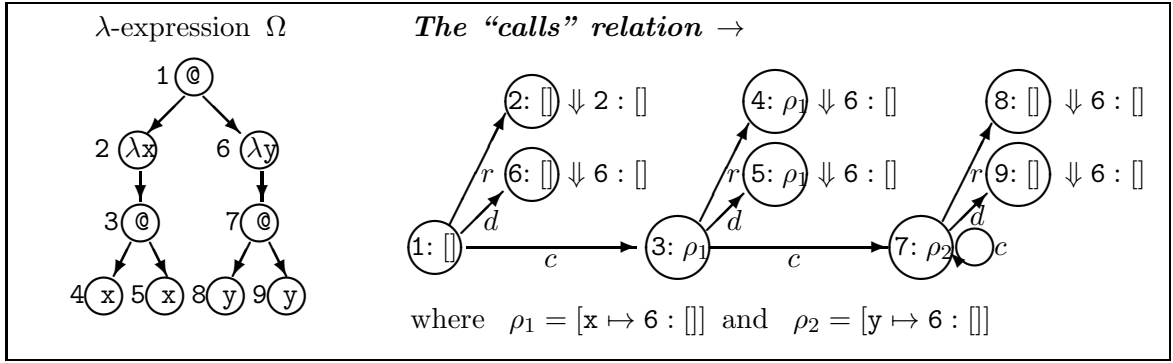
\begin{fig}{The DCG or dynamic control graph of a $\lambda$-expression}{htb}{fig-ex-call relation}

    \fl{\tt    \setlength{\unitlength}{0.92pt}
\begin{picture}(530,140)(60,40)
\thicklines

              \put(70,170){$\lambda${\rm-expression} $\Omega$}

              \put(95,150){1 @}
              \put(111,153){\circle{15}}
              \put(107,145){\vector(-1,-1){15}}
              \put(115,145){\vector(1,-1){15}}
	      
              \put(70,120){2 $\lambda$x}
              \put(88,123){\circle{15}}
              \put(89,114){\vector(0,-1){12}}
	      
              \put(120,120){6 $\lambda$y}
              \put(138,123){\circle{15}}
              \put(139,114){\vector(0,-1){12}}
	      
              \put(74,90){3 @}
              \put(89,93){\circle{15}}
              \put(85,85){\vector(-1,-2){7}}
              \put(93,85){\vector(1,-2){7}}
	      
              \put(60,60){4 x}
              \put(74,63){\circle{15}}
	      
              \put(85,60){5 x}
              \put(99,63){\circle{15}}
	      
              \put(124,90){7 @}
              \put(138,93){\circle{15}}
              \put(134,85){\vector(-1,-2){7}}
              \put(142,85){\vector(1,-2){7}}
	      
              \put(110,60){8 y}
              \put(124,63){\circle{15}}
	      
              \put(135,60){9 y}
              \put(149,63){\circle{15}}

              \put(200,170){{\bf\em The ``calls'' relation} $\to$}
	      
              \put(200,80){1$:[]$}
              \put(210,83){\circle{20}}
              \put(223,83){\vector(1,0){70}}
              \put(255,75){$c$}
              \put(221,89){\vector(1,1){17}}
              \put(229,110){$r$}
              \put(216,92){\vector(1,2){23}}
              \put(230,90){$d$}
	      
              \put(240,110){6$:[]\,\Downarrow{\tt 6}:[]$}
              \put(250,113){\circle{20}}
	      
              \put(240,140){2$:[]\,\Downarrow{\tt 2}:[]$}
              \put(250,143){\circle{20}}
	      
              \put(299,80){3$:\rho_1$}
              \put(310,83){\circle{24}}
              \put(323,83){\vector(1,0){73}}
              \put(355,75){$c$}
              \put(322,90){\vector(1,1){15}}
              \put(330,110){$r$}
              \put(319,94){\vector(1,2){20}}
              \put(330,90){$d$}
	      
              \put(340,110){5$:\rho_1 \Downarrow{\tt 6}:[]$}
              \put(350,113){\circle{24}}
	      
              \put(340,140){4$:\rho_1\Downarrow{\tt 6}:[]$}
              \put(350,143){\circle{24}}
	      
              \put(399,80){7$:\rho_2$}
              \put(410,83){\circle{24}}
              \put(422,90){\vector(1,1){15}}
              \put(430,110){$r$}
              \put(419,94){\vector(1,2){20}}
              \put(430,92){$d$}
              \put(432,83){\circle{15}}
              \put(425,78){\vector(-1,1){3}}
              \put(440,80){$c$}
	      
              \put(440,110){9$:[]\  \Downarrow{\tt 6}:[]$}
              \put(450,113){\circle{24}}
	      
              \put(440,140){8$:[]\  \Downarrow{\tt 6}:[]$}
              \put(450,143){\circle{24}}

              \put(200,50){{\rm where \hair
	      $\rho_1=[{\tt x}\mapsto{\tt 6}:[]]$\hair and \hair
	      $\rho_2=[{\tt y}\mapsto{\tt 6}:[]]$}}
\end{picture}}

\end{fig}

\fl By the NIS Lemma \ref{lem-NIS-environment}, 
if  ${\tt P}\not\Downarrow$ then there  exists
an infinite call chain
$$
{\tt P}:[] = {\tt e}_0:\rho_0\to{\tt e}_1:\rho_1\to{\tt 
e}_2:\rho_2\to \ldots 
$$

\fl By Lemma \ref{lem-subexpression}, 
${\tt e}_i\in{\it subexp}({\tt P})$ for each $i$. 
Our termination-detecting algorithm will focus on the {\em size relations 
between consecutive environments} $\rho_i$ and $\rho_{i+1}$ in this chain.
Since ${\it subexp}({\tt P})$ is a finite set,  at least 
one subexpression {\tt e} occurs 
infinitely often, so ``self-loops'' will be of particular 
interest.

Since all states have an expression component lying in a set of fixed 
size, 
and each expression in the environment also lies in this finite set,
in an infinite state set $\cal S$
there will be states whose {\em environment depths} are arbitrarily large.

\subsection{Static control flow graphs for $\lambda$-expressions} 
\label{sec-abstract-interpretation-for-cfg}

The end goal, given program {\tt P}, is implied by the NIS Lemma 
\ref{lem-NIS-environment}:
correctly to assert the 
nonexistence of any infinite call chain starting at ${\tt P}:[]$.
By  the Subexpression Lemma 
\ref{lem-subexpression} an  infinite  call chain 
${\tt e}_0:\rho_0\to{\tt e}_1:\rho_1\to{\tt 
e}_2:\rho_2\to \ldots$
can only contain  
finitely many different expression components ${\tt e}_i$.
A static control flow graph (SCG for short) including all expression components can be
obtained by abstract interpretation
 of the ``Calls'' and  ``Evaluates-to'' relations
  (Cousot and Cousot  \cite{CousotCousot}). 
  Figure \ref{fig-control-flow-graph} shows a SCG for $\Omega$.
  
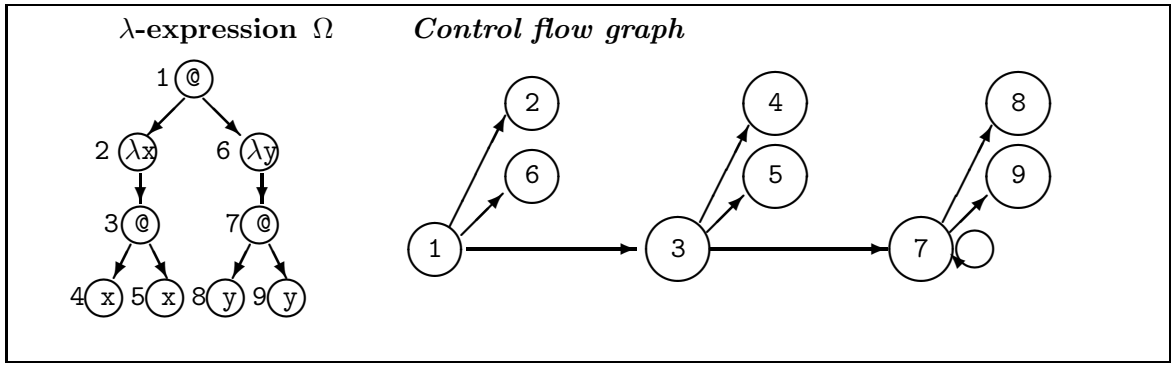
  \begin{fig}{The SCG or static control graph of a $\lambda$-expression}{htb}
	       {fig-control-flow-graph}

    \fl{\tt    \setlength{\unitlength}{0.92pt}
\begin{picture}(530,140)(60,40)
\thicklines

              \put(80,170){$\lambda${\bf-expression} $\Omega$}

              \put(95,150){1 @}
              \put(111,153){\circle{15}}
              \put(107,145){\vector(-1,-1){15}}
              \put(115,145){\vector(1,-1){15}}
	      
              \put(70,120){2 $\lambda$x}
              \put(88,123){\circle{15}}
              \put(89,114){\vector(0,-1){12}}
	      
              \put(120,120){6 $\lambda$y}
              \put(138,123){\circle{15}}
              \put(139,114){\vector(0,-1){12}}
	      
              \put(74,90){3 @}
              \put(89,93){\circle{15}}
              \put(85,85){\vector(-1,-2){7}}
              \put(93,85){\vector(1,-2){7}}
	      
              \put(60,60){4 x}
              \put(74,63){\circle{15}}
	      
              \put(85,60){5 x}
              \put(99,63){\circle{15}}
	      
              \put(124,90){7 @}
              \put(138,93){\circle{15}}
              \put(134,85){\vector(-1,-2){7}}
              \put(142,85){\vector(1,-2){7}}
	      
              \put(110,60){8 y}
              \put(124,63){\circle{15}}
	      
              \put(135,60){9 y}
              \put(149,63){\circle{15}}

              \put(200,170){{\bf\em Control flow graph}}
	      
              \put(207,80){1}
              \put(210,83){\circle{20}}
              \put(223,83){\vector(1,0){70}}
              \put(221,89){\vector(1,1){17}}
              \put(216,92){\vector(1,2){23}}	      
              \put(247,110){6}
              \put(250,113){\circle{20}}
	      
              \put(247,140){2}
              \put(250,143){\circle{20}}
	      
              \put(307,80){3}
              \put(310,83){\circle{24}}
              \put(323,83){\vector(1,0){73}}
              \put(322,90){\vector(1,1){15}}
              \put(319,94){\vector(1,2){20}}	      
              \put(347,110){5}
              \put(350,113){\circle{24}}
	      
              \put(347,140){4}
              \put(350,143){\circle{24}}
	      
              \put(407,80){7}
              \put(410,83){\circle{24}}
              \put(422,90){\vector(1,1){15}}
              \put(419,94){\vector(1,2){20}}
              \put(432,83){\circle{15}}
              \put(425,78){\vector(-1,1){3}}	      
              \put(447,110){9}
              \put(450,113){\circle{24}}
	      
              \put(447,140){8}
              \put(450,143){\circle{24}}

\end{picture}}

\end{fig}

An approximating SCG may be obtained by removing all 
environment components from 
Definition \ref{def-evaluate-and-call-relation-environment}. To deal with the 
absence of environments the variable lookup rule 
is modified: 
If ${\tt e}_1{\tt@e}_2$ is {\em any} application in {\tt P} 
such that ${\tt e}_1$ can evaluate to a value of form $\lambda{\tt 
x.e}$ and 
${\tt e}_2$ can evaluate to value $v_2$, then $v_2$
is regarded as a possible value of {\tt x}. 

Although approximate, these rules have the virtue that 
there are only finitely many possible judgements ${\tt e}\to{\tt e}'$ 
and
${\tt e}\Downarrow{\tt e}'$. Consequently, the runtime behavior of 
program {\tt P} may be 
(approximately) analysed by exhaustively applying these 
inference rules. A later section will extend the rules so they also 
generate  size-change graphs.

\bdfn
\label{def-eval-approx}
{\rm(Approximate evaluation and call rules)}
The new judgement forms are
${\tt e}\Downarrow{\tt e}'$ and ${\tt e}\to{\tt e}'$. 
The inference rules are:
\smallskip
$$
\infer[(\textnormal{ValueA})]{\lambda{\tt x.e} \Downarrow
  \lambda{\tt x.e}}{}
\qquad
\infer[(\textnormal{VarA})]{{\tt x}\Downarrow v_2}
{{\tt e}_1{\tt@e}_2\in{\it subexp}({\tt P})
  &
  {\tt e}_1\Downarrow\lambda{\tt x.e}_0
  &
   {\tt e}_2\Downarrow v_2}
$$

$$
\infer[(\textnormal{OperatorA})]{{\tt e}_1{\tt@e}_2\tosub{r}{\tt e}_1}
{}
\qquad
\infer[(\textnormal{OperandA})]{{\tt e}_1{\tt@e}_2\tosub{d}{\tt e}_2}
{}
$$

$$
\infer[(\textnormal{CallA})]{{\tt e}_1{\tt@e}_2\tosub{c}{\tt e}_0}
{{\tt e}_1\Downarrow\lambda{\tt x.e}_0 \hair {\tt e}_2\Downarrow v_2}
\qquad
\infer[(\textnormal{ApplyA})]{{\tt e}_1{\tt@e}_2\Downarrow v}
{{\tt e}_1{\tt@e}_2\tosub{c}{\tt e}'
  & 
  {\tt e}'\Downarrow v}
$$
\edfn

\fl  The (VarA) rule refers globally to {\tt P}, the program being 
analysed.
The approximate evaluation is nondeterministic, since an 
expression may evaluate to more than one value. 
\medskip

Following is a central result: that 
all possible values obtained by the {\em actual evaluation} of  Definition \ref{def-evaluate-and-call-relation-environment}
are accounted for by
the {\em approximate evaluation} of Definition \ref{def-eval-approx}.
\blem \label{abs-super} 
\hfill  \bt{ccccc} If ${\tt P}:[]\to^*{\tt e}:\rho$ &and& ${\tt e}:\rho\Downarrow{\tt e}':\rho'$, &then& 
${\tt e}\Downarrow{\tt e}'$. \\

\hfill  If  ${\tt P}:[]\to^*{\tt e}:\rho$  &and& ${\tt e}:\rho\to{\tt e}':\rho'$, &then& ${\tt e}\to{\tt e}'$.
\et
\elem
\fl Proof is in the Appendix.
\section{A quick review of size-change analysis}
 \label{sec-quick-review-size-change-analysis}

Using the framework of  \cite{popl01}, the relation between two
states $s_1$ and $s_2$ in a call 
$s_1\to s_2$ or an evaluation $s_1\Downarrow s_2$ will be
described by means of a size-change graph $G$.

\begin{exa}
Let first-order  functions {\tt f} and {\tt g} be
defined by mutual recursion:

\bp
    f(x,y)   = if x=0 then y else {1:} g(x,y,y) 
    g(u,v,w) = if w=0 then {3:}f(u-1,w) else {2:}g(u,v-1,w+2)

\ep

\fl Label the three function calls  {\tt 1}, {\tt 2} and {\tt 
3}. The ``control flow graph'' in Figure \ref{fig-first-order-example} 
shows the calling function
and called function of each 
call, e.g., ${\tt 1}:{\tt f}\to{\tt g}$. Associate with each call 
a ``size-change graph'', e.g., $G_1$ for call {\tt 1}, that safely describes 
the data flow from the calling function's parameters to the 
called function's parameters. Symbol $\downarrow$ indicates a value decrease.
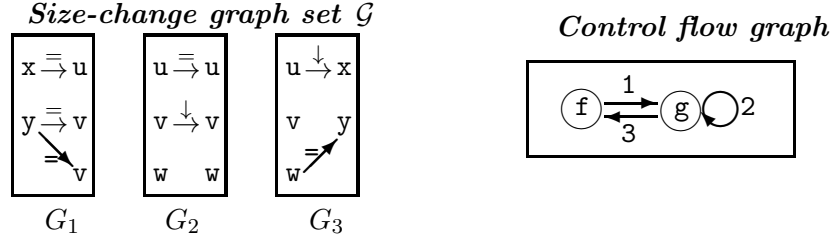
\begin{fig0}{Call graph and size-change graphs for the example 
first-order program.}{htb}
	       {fig-first-order-example}
	       
\fl{\tt    \setlength{\unitlength}{0.5pt}
\begin{picture}(530,180)(-100,20)
\thicklines

              \put(60,190){{\bf\em Size-change graph set} $\cal G$}

              \put(50,60){\framebox(60,120){}}
              \put(56,150){x}
              \put(95,150){u}
              \put(56,110){y}
              \put(95,110){v}
              \put(95,70){v}
              \put(74,34){$G_1$}
              \put(70,148){$\stackrel{=}{\to}$}
              \put(70,107){$\stackrel{=}{\to}$}
              \put(70,107){\vector(1,-1){25}}
              \put(73,80){=}

              \put(150,60){\framebox(60,120){}}
              \put(156,150){u}
              \put(195,150){u}
              \put(156,110){v}
              \put(195,110){v}
              \put(156,70){w}
              \put(195,70){w}
              \put(165,34){$G_2$}
              \put(170,107){$\stackrel{\downarrow}{\to}$}
              \put(170,148){$\stackrel{=}{\to}$}

              \put(250,60){\framebox(60,120){}}
              \put(256,150){u}
              \put(295,150){x}
              \put(256,110){v}
              \put(295,110){y}
              \put(256,70){w}
              \put(270,77){\vector(1,1){25}}
              \put(270,85){=}
              \put(274,34){$G_3$}
              \put(270,148){$\stackrel{\downarrow}{\to}$}

              \put(460,180){{\bf\em Control flow graph}}
	      
              \put(440,90){\framebox(200,70)}

              \put(475,120){f}
              \put(550,120){g}
              \put(497,129){\vector(1,0){40}}
              \put(537,119){\vector(-1,0){40}}

              \put(600,120){2}
              \put(510,137){1}
              \put(510,100){3}
              \put(585,125){\circle{25}}
              \put(573,120){\vector(-1,1){3}}
	      \thinlines
              \put(480,125){\circle{30}}
              \put(555,123){\circle{30}}

\end{picture}}

\end{fig0}

\fl{\bf Termination reasoning:}
We show that
{\em all infinite size-change graph sequences}  
${\cal M}= g_1g_2\ldots \in \{G_1,G_2,G_3\}^\omega$
that follow 
the program's control flow {\em are impossible} (assuming
that the data value set is well-founded):
\vair

{\bf Case 1:}\ \    ${\cal M} \in \ldots(G_2)^{\omega}$ ends in infinitely many $G_2$'s: 
This would imply that {\em variable {\tt v} descends infinitely.}

{\bf Case 2:}\ \  ${\cal M} \in \ldots(G_1G_2^*G_3)^{\omega}$. 
This would imply that  {\em variable {\tt u} descends infinitely.}

\fl Both cases are impossible; therefore a call of  
{\em any program function with any data will terminate.}
\hfill {\em End of example.}

\end{exa}

\bdfn
\label{def-size-change-terminology}
\hfill

\be[(1)]
	   
\item 

A {\em size-change graph} $A\stackrel{G}{\to}B$ consists of 
a {\em source} set $A$; 
a {\em target} set $B$; and 
a   set of labeled\footnote{Arc label $\deq$ signifying $\geq$ was used in  \cite{popl01} instead of $=$, but this makes no difference in our context.}
 arcs $G\subseteq A \times \{=,\downarrow\}\times B$.

\item The {\em identity}  size-change graph for $A$ is 
$A\stackrel{{\it id}_A}{\to}A$ where 
${\it id}_A = \{{\tt x}\stackrel{=}{\to}{\tt x}\ |\ {\tt x}\in A\}$.

\item Size-change graphs $A\stackrel{G_1}{\to}B$ and 
$C\stackrel{G_2}{\to}D$ are {\em composible} if $B=C$.
The {\em composition} of 
$A\stackrel{G_1}{\to}B$ and 
$B\stackrel{G_2}{\to}C$  is
$A\stackrel{G_1;G_2}{\longrightarrow}C$
where 


\ee

$
\ba{rl} 
\quad G_1;G_2 = & \{{\tt x}\stackrel{\downarrow}{\to}{\tt z}\ |\ 
 \ \downarrow\ \ \ \in  \{\,r,s\ |\  
   {\tt x}\stackrel{r}{\to}{\tt y}\in G_1\mbox{\ and\ }
   {\tt y}\stackrel{s}{\to}{\tt z}\in G_2\mbox{\ for some\ }{\tt y}\in B\}\ \} \\

\cup &  \{{\tt x}\stackrel{=}{\to}{\tt z}\ |\ 
 \{=\} = \{\,r,s\ |\  
   {\tt x}\stackrel{r}{\to}{\tt y}\in G_1\mbox{\ and\ }
   {\tt y}\stackrel{s}{\to}{\tt z}\in G_2\mbox{\ for some\ }{\tt y}\in B\}\ \}
\ea
$
\edfn

\blem Composition is associative, and $A\stackrel{G}{\to}B$
implies ${\it id}_A;G = G;{\it id}_B = G$.
\elem

\bdfn A {\em multipath} ${\cal M}$ over a set $\cal G$  of 
size-change graphs is a finite or infinite composible sequence
of graphs in $\cal G$. Define
$${\cal G}^{\omega} = \{{\cal M}=G_0, G_1,\ldots \  |  \ 
        \mbox{graphs $G_i,G_{i+1}$ are composible for\ } 
	i= 0,1,2,\ldots\ \}$$ 
\edfn

\bdfn\label{def-thread-inf-descent}\hfill
\be[(1)]
\item A {\em thread} in a multipath 
${\cal M} = G_0, G_1, G_2,\ldots$ 
is a sequence 
$t = a_j \stackrel{r_j}{\to} a_{j+1} \stackrel{r_{j+1}}{\to} \ldots$ 
such that 
$a_k \stackrel{r_k}{\to} a_{k+1}\in G_k$ for every $k\geq j$
(and each $r_k$ is $=$ or $\downarrow$.)\vair

\item Thread $t$ is of {\em infinite descent} if $r_k =\, \downarrow$ for 
infinitely many $k\geq j$.\vair
\ee
\edfn

\bdfn \label{def-size-change-condition}The size-change condition.\hfill
\be
\item[] A  set  $\cal G$ of size-change graphs satisfies the 
{\em size-change condition} if every infinite multipath 
${\cal M}\in {\cal G}^{\omega}$ 
contains at least one thread of infinite descent.
\ee
\edfn

\fl Perhaps surprisingly, the size-change condition is decidable. 
Its worst-case complexity
is shown to be complete for {\sc pspace}  in \cite{popl01}
(for first-order programs, in relation to the length of the program being analysed).
\smallskip

\fl{\it The example revisited} 
 The program of Figure \ref{fig-first-order-example} 
has three size-change graphs, one for each of the calls 
${\tt 1}:{\tt f}\to{\tt g}, 
 {\tt 2}:{\tt g}\to{\tt g},
 {\tt 3}:{\tt g}\to{\tt f}$, so
 ${\cal G}=\{A\stackrel{G_1}{\to}B,B\stackrel{G_2}{\to}B,
 B\stackrel{G_3}{\to}A\}
 $
 where $A= \{{\tt x},{\tt y}\}$ and $B=\{{\tt u},{\tt v},{\tt w}\}$.
(Note: the vertical layout
of size-change graphs in Figure \ref{fig-first-order-example} is 
inessential; one could simply write
$G_3 =\{ {\tt u}\stackrel{\downarrow}{\to}{\tt x}, 
         {\tt w}\stackrel{=}{\to}{\tt y}\}$.)

	${\cal G}$  
satisfies the size-change condition, since {\em every infinite multipath 
has either a thread that decreases {\tt u} infinitely, or
a thread that decreases {\tt v} infinitely}.

\section{Tracing data size changes in call-by-value $\lambda$-calculus evaluation}

The next focus is on size relations between consecutive environments 
in a call chain.

\subsection{Size changes in a computation: a well-founded  relation between states}

\bdfn 
\label{def-graph-nodes}\hfill
\be[(1)]

\item A {\em name path} is  a finite string $p$ of 
variable names, where the empty string is (as usual) written $\epsilon$. 

\item The {\em graph basis} of a state $s={\tt e}:\rho$ is the smallest set  
${\it gb}(s)$ 
of name paths satisfying
$$
{\it gb}({\tt e:}\rho) = \{\epsilon\}\cup
\{ {\tt x}p\ |\ 
{\tt x}\in {\it fv}(\tt e) 
\mbox{\ and\ }
p\in{\it gb}(\rho({\tt x}))\}
$$
\ee
\edfn
By this definition, for the two states in the example above we have
$
{\it gb}(s) = \{\epsilon,{\tt r},{\tt a}\}$ and 
${\it gb}(s') = \{\epsilon,{\tt r},{\tt r}{\tt r},{\tt a}\}
$. Further, given a state $s$ and a path $p\in{\it gb}(s)$, we can 
find the substate identified by name path $p$ as follows: 

\bdfn 
\label{def-graph-valuation}
The {\em valuation function}
$\overline{s}:{\it gb}(s)\to{\it State}$  of a state $s$ 
is defined by: 
 $$\overline{s}(\epsilon)=s\mbox{\rm\hair and \hair}
 \overline{{\tt e}:\rho}({\tt x}p)=\overline{\rho({\tt x})}(p)$$
\edfn
We need to develop a size ordering on states. 
This will  be modeled by size-change arcs
$\stackrel{=}{\to}$ and $\stackrel{\downarrow}{\to}$.
The size relation we use is partly the ``subtree'' relation on closure 
values ${\tt e}:\rho$, and partly the ``subexpression'' relation on 
$\lambda$-expressions.

\bdfn\label{def-value-decrease}\hfill

\be[(1)]

\item 
The {\em state support}  of 
a state ${\tt e}:\rho$ is given by
\[
{\it support}({\tt e}:\rho) = \{{\tt e}:\rho\}\  \cup 
 \bigcup_{{\tt x}\in{\it fv}({\tt e})}
     {\it support}(\rho({\tt x}))
     \]

\item Relations  
$\succ_1$, $\succ_2$, $\succeq$ and $\succ$ on states are defined by:
\be[$\bullet$]

\item  $s_1\succ_1 s_2$ holds if ${\it support}(s_1)\ni s_2$ and $s_1 \neq s_2$; 

\item $s_1\succ_2 s_2$ holds if $s_1={\tt e}_1:\rho_1$ and 
$s_2={\tt e}_2:\rho_2$, where 
$ {\it subexp}({\tt e}_1) \ni {\tt e}_2$ and ${\tt e_1}\neq {\tt e}_2$ and $\forall {\tt x}\in{\it fv}({\tt e}_2).\rho_1(x)=\rho_2(x)$. Further, \\

\item Relation  $\succeq$ is defined to be the transitive closure of 
$\succ_1\cup \succ_2\cup =$. 

\item Finally, $s_1 \succ s_2$ if $s_1\succeq s_2$ and $s_1\neq s_2$\\
\ee
\ee
\edfn

\blem\label{lem-well-founded}
The relation $\succ\ \subseteq{\it State}\times{\it State}$ is 
well-founded.
\elem

We prove that the relation $\succ$ on states is well-founded
by proving that 
$${\tt e}_1:\rho_1 \succ {\tt e}_2:\rho_2 \mbox{ implies  
that } (H({\tt e}_1:\rho_1),L({\tt e}_1)) >_{lex} (H({\tt e}_2:\rho_2),L({\tt e}_2))$$
in the lexicographic order, where $H$ gives the height of the environment and $L$ gives the length of the expression. 
 The proof is in the Appendix.

\blem\label{lem-sub-env}
If $p\in gb(s)$ then $s \succeq_1 \overline{s}(p)$. 
If $p\in gb(s)$ and $p\neq \epsilon$ then $s \succ_1 \overline{s}(p)$.
\elem

\section{Size-change graphs that safely describe a program}

\subsection{Safely describing state transitions}

We now define the arcs of the size-change graphs 
(recalling Definition \ref{def-size-change-terminology}):
\bdfn 
\label{def-graph-arcs}
A size-change graph $G$ relating state $s_1$ to state $s_2$ has {\em source}
${\it gb}(s_1)$ and {\em target}
${\it gb}(s_2)$. 
\edfn

\bdfn\label{def-safe-call-and-evaluation}
Let
$s_1 = {\tt e}_1:\rho_1$ and $s_2 = {\tt e}_2:\rho_2$.
Size-change graph ${\tt s}_1{\to}{\tt s}_2,G$ is 
{\em safe}\footnote{The term ``safe'' comes from abstract interpretation 
\cite{jonesnielson}. An alternative would be  ``sound.''}  
for $(s_1, s_2)$ if
$$
p_1\stackrel{=}{\to}p_2 \in G\mbox{\rm\ \ implies\ \ }
\overline{s_1}(p_1) = \overline{s_2}(p_2){\rm\ \ and\ \ }
p_1\stackrel{\downarrow}{\to}p_2 \in G\mbox{\rm\ \ implies\ \ }
\overline{s_1}(p_1) \succ \overline{s_2}(p_2)
$$
\edfn
\fl
By $dom(G)$ we denote the subset of $source(G)$ from where arcs begin. By $codom(G)$ we denote the subset of $target(G)$ where arcs end.
Notice that if a size-change graph $G$ is safe for the states $(s_1,s_2)$, then any subset size-change graph $G'\subset G$ with $source(G')= source(G)$ and $target(G') = target(G)$ is safe for $(s_1,s_2)$.

\bdfn
\label{safe-set-of-size-change-graphs}
A set $\cal G$ of size-change graphs is 
{\em safe for program {\tt P}} if  
$\ {\tt P}:[]\ \to^*\ s_1\to s_2$ implies some 
$G\in\cal G$ is safe 
for the pair $(s_1,s_2)$.
\edfn

\begin{exa} Figure \ref{fig-ann-call-graph} below shows a graph 
set $\cal G$ that is safe for the program 
$\Omega = (\lambda${\tt x.x@x}$)(\lambda${\tt y.y@y}$)$.
For brevity, each subexpression of $\Omega$  is 
referred to by number in the diagram of $\cal G$.
Subexpression ${\tt 1} = \Omega $ has no free 
variables, so  arcs from node {\tt 1} are labeled with
size-change graphs $G_0 = \emptyset$.
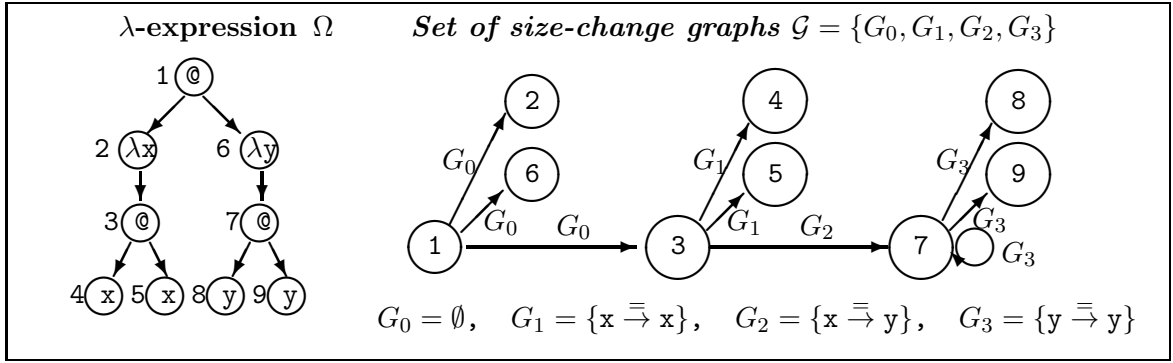
\begin{fig}{A set of size-change graphs that safely describe
     $\Omega$'s nonterminating computation}{htb}
	       {fig-ann-call-graph}

    \fl{\tt    \setlength{\unitlength}{0.92pt}
\begin{picture}(530,140)(60,40)
\thicklines

              \put(80,170){$\lambda${\bf-expression} $\Omega$}

              \put(95,150){1 @}
              \put(111,153){\circle{15}}
              \put(107,145){\vector(-1,-1){15}}
              \put(115,145){\vector(1,-1){15}}
	      
              \put(70,120){2 $\lambda$x}
              \put(88,123){\circle{15}}
              \put(89,114){\vector(0,-1){12}}
	      
              \put(120,120){6 $\lambda$y}
              \put(138,123){\circle{15}}
              \put(139,114){\vector(0,-1){12}}
	      
              \put(74,90){3 @}
              \put(89,93){\circle{15}}
              \put(85,85){\vector(-1,-2){7}}
              \put(93,85){\vector(1,-2){7}}
	      
              \put(60,60){4 x}
              \put(74,63){\circle{15}}
	      
              \put(85,60){5 x}
              \put(99,63){\circle{15}}
	      
              \put(124,90){7 @}
              \put(138,93){\circle{15}}
              \put(134,85){\vector(-1,-2){7}}
              \put(142,85){\vector(1,-2){7}}
	      
              \put(110,60){8 y}
              \put(124,63){\circle{15}}
	      
              \put(135,60){9 y}
              \put(149,63){\circle{15}}

              \put(200,170){{\bf\em Set of size-change graphs 
	                     ${\cal G}=\{G_0,G_1,G_2,G_3\}$}}
	      
              \put(207,80){1}
              \put(210,83){\circle{20}}
              \put(223,83){\vector(1,0){70}}
              \put(221,89){\vector(1,1){17}}
              \put(216,92){\vector(1,2){23}}
              \put(213,115){$G_0$}
              \put(230,90){$G_0$}
              \put(260,88){$G_0$}
	      
              \put(247,110){6}
              \put(250,113){\circle{20}}
	      
              \put(247,140){2}
              \put(250,143){\circle{20}}
	      
              \put(307,80){3}
              \put(310,83){\circle{24}}
              \put(323,83){\vector(1,0){73}}
              \put(322,90){\vector(1,1){15}}
              \put(319,94){\vector(1,2){20}}
              \put(316,115){$G_1$}
              \put(330,90){$G_1$}
              \put(360,88){$G_2$}
	      
              \put(347,110){5}
              \put(350,113){\circle{24}}
	      
              \put(347,140){4}
              \put(350,143){\circle{24}}
	      
              \put(407,80){7}
              \put(410,83){\circle{24}}
              \put(422,90){\vector(1,1){15}}
              \put(419,94){\vector(1,2){20}}
              \put(432,83){\circle{15}}
              \put(425,78){\vector(-1,1){3}}
              \put(416,115){$G_3$}
              \put(431,91){$G_3$}
              \put(443,77){$G_3$}
	      
              \put(447,110){9}
              \put(450,113){\circle{24}}
	      
              \put(447,140){8}
              \put(450,143){\circle{24}}

              \put(180,50){
	      $G_0 = \emptyset$, \
	      $G_1 = \{ {\tt x}\stackrel{=}{\to}{\tt x} \}$, \ 
	      $G_2 = \{ {\tt x}\stackrel{=}{\to}{\tt y} \}$, \ 
	      $G_3 = \{ {\tt y}\stackrel{=}{\to}{\tt y} \}$ } 

\end{picture}}
\end{fig}
\end{exa} 

\bthm
If $\cal G$ is safe for program {\tt P} and satisfies 
the size-change condition, then call-by-value evaluation of {\tt P} 
terminates.
\ethm

\bprf
Suppose call-by-value-evaluation of {\tt P} does not terminate. Then by 
Lemma \ref{lem-NIS-environment} there is an  infinite call chain
$$
{\tt P}:[] = {\tt e}_0:\rho_0\to{\tt e}_1:\rho_1\to{\tt 
e}_2:\rho_2\to \ldots 
$$
Letting $s_i={\tt e}_i:\rho_i$, by safety of $\cal G$ 
(Definition \ref{safe-set-of-size-change-graphs}),  
there is a size-change 
graph $G_i\in\cal G$ that safely describes each pair
$(s_i,s_{i+1})$.
By the size-change condition 
(Definition \ref{def-size-change-condition}) the multipath 
${\cal M}=G_0,G_1,\ldots$ has an infinite  thread
$t = a_j \stackrel{r_j}{\to} a_{j+1} \stackrel{r_{j+1}}{\to} \ldots$ 
such that $k\geq j$ implies
$a_k \stackrel{r_k}{\to} a_{k+1}\in G_k$, and each $r_k$ is 
$\downarrow$ or $=$, and 
there are infinitely many $r_k = \,\downarrow$.
Consider the value sequence 
$\overline{s_j}(a_j), \overline{s_{j+1}}(a_{j+1}),\ldots$.
By safety of $G_k$  (Definition \ref{def-safe-call-and-evaluation})
we have 
$\overline{s_k}(a_k)\succeq\overline{s_{k+1}}(a_{k+1})$  
for every $k\geq j$,
and infinitely many  
proper decreases 
$\overline{s_k}(a_k)\succ\overline{s_{k+1}}(a_{k+1})$. 
However this 
is impossible since by Lemma \ref{lem-well-founded} the 
relation $\succ$ on {\it State} is 
well-founded.

Conclusion: call-by-value-evaluation of {\tt P} 
terminates.
\eprf

\fl The goal is partly achieved: We have found a sufficient condition 
on a set of size-change graphs to guarantee program termination. 
What we have not yet done is to find an algorithm to 
{\em construct} a size-change graph set $\cal G$ that is safe for {\tt P}
(The safety condition of 
Definition \ref{safe-set-of-size-change-graphs} is in general undecidable, so
enumeration of all graphs won't work.)
Our graph construction algorithm is developed in two stages:
\be[$\bullet$]
\item 
First, the exact evaluation and call relations are ``instrumented'' so 
as to produce safe size-change graphs during evaluation. 

\item Second, an extension of the abstract interpretation 
from Section \ref{sec-abstract-interpretation-for-cfg} yields a 
{\em computable} over-approximation $\cal G$ that  contains all graphs that can be 
built during exact evaluation. 
\ee

\subsection{Generating size-change graphs during a computation}
We now ``instrument''  the exact evaluation and call relations  so 
as to produce safe size-change graphs during evaluation. In the definition of the size-change graphs {\tt x}, {\tt y}, {\tt z} are variables, and  $p,q$ can be variables or $\epsilon$, the empty path. Recall the valuation function for a state gives $\bar s(\epsilon)=s$, so in a sense $\epsilon$ is bound to the whole state. 

\bdfn
\label{def-eval-call-generate-graphs}
{\rm(Evaluation and call with graph generation)}
The extended  evaluation and call judgement forms are 
${\tt e}:\rho\to {\tt e}':\rho', G
\mbox{\hair\hair and\hair\hair}
{\tt e}:\rho\Downarrow{\tt e}':\rho', G$, where $source(G)$ = fv({\tt e})$\cup\{\epsilon\}$ and $target(G)$ = fv({\tt e}$'$)$\cup\{\epsilon\}$. The inference rules are:
\medskip
$$
\infer[(\textnormal{ValueG})]{\lambda{\tt x.e}:\rho \Downarrow
  \lambda{\tt x.e}:\rho,{\it id}^=_{\lambda{\tt x.e}}}
{}
$$
$$
\infer[(\textnormal{OperatorG})]{{\tt e}_1{\tt@e}_2:\rho\tosub{r}{\tt e}_1:\rho,
   {\it id}^\downarrow_{{\tt e}_1}}
{}
\qquad\qquad
\infer[(\textnormal{OperandG})]{{\tt e}_1{\tt@e}_2:\rho
\tosub{d}
{\tt e}_2:\rho, 
   {\it id}^\downarrow_{{\tt e}_2}}
{{\tt e}_1:\rho\Downarrow v_1}
$$
\smallskip

\fl\bt{cll}
${\it id}^=_e$ &stands for& 
$\{\epsilon\stackrel{=}{\to} \epsilon\}\cup \{{\tt y}\stackrel{=}{\to}{\tt y}\  |\ {\tt y} \in fv({\tt e})\}$ \\
${\it id}^\downarrow_e$ &stands for& 
$\{\epsilon\stackrel{\downarrow}{\to}\epsilon\}\cup
 \{{\tt y}\stackrel{=}{\to}{\tt y}\ |\  {\tt y} \in fv({\tt e})\}$ \\
\et\\
\smallskip

\noindent An arc ${\tt y}\stackrel{=}{\to}{\tt y}$ express that the
state bound to the variable ${\tt y}$ is the same in both sides,
before and after the evaluation or call.\\ The $\epsilon$ ``represent"
the whole state.  In the (ValueG) rule the state $\lambda{\tt
x.e}:\rho$ is the same in both sides and so there is an arc
$\epsilon\stackrel{=}{\to} \epsilon$. In the (OperatorG) and
(OperandG) rules the state is smaller in the right hand side because
we go to a strict subexpression and possibly also restrict the
environment $\rho$ accordingly. So there are
$\epsilon\stackrel{\downarrow}{\to}\epsilon$ arcs.

$$
\infer[\mbox{{\it {\small $\rho({\tt x})={\tt e}':\rho'$}}\ }(\textnormal{VarG})]{{\tt x}:\rho\Downarrow \rho({\tt x}), 
 \ \{{\tt x}\stackrel{\downarrow}{\to}{\tt y}\ |{\tt y}\in fv({\tt e}')\ \}
\cup\{{\tt x}\stackrel{=}{\to}\epsilon\}}
{}
$$
\smallskip

\noindent In the (VarG) rule the state on the right side is $\rho(x)$. This is the state which $x$ is bound to in the environment in the left hand side, therefore we have an arc ${\tt x}\stackrel{=}{\to}\epsilon$. Suppose $\rho({\tt x})={\tt e}':\rho'$ and ${\tt y}\in fv({\tt e}')$. Then ${\tt y}$ is bound in $\rho'$ and this binding is then a subtree of  ${\tt e}':\rho'$. So we have an arc ${\tt x}\stackrel{\downarrow}{\to}{\tt y}$.

$$
\infer[(\textnormal{CallG})]{{\tt e}_1{\tt@e}_2:\rho\tosub{c}{\tt e}_0:\rho_0[{\tt x}\mapsto v_2],
G_1^{-\epsilon/\lambda x.e_0}\cup_{e_0}  G_2^{\epsilon\mapsto{\tt x}}}
{{\tt e}_1:\rho\Downarrow \lambda{\tt x.e}_0:\rho_0,G_1
 & {\tt e}_2:\rho\Downarrow v_2,G_2}
$$

\noindent In the definition of the size-change graphs used in the
(CallG) rule {\tt x}, {\tt y}, {\tt z} are variables, and $p,q$ can be
variables or $\epsilon$.  In $\stackrel{r}{\to}$ the $r$ can be either
$\downarrow$ or $=$.  The construction of the size-change graph
associated with the call is explained below.  \vair

\fl\bt{lll}
$G_1^{-\epsilon/\lambda x.e_0}$&stands for&cases\\

&${\tt x}\in fv({\tt e}_0)$&:   
$\{{\tt y}\stackrel{r}{\to}{\tt z}  \ |\  {\tt y}\stackrel{r}{\to}{\tt z} \in G_1\}\cup
\{\epsilon\stackrel{\downarrow}{\to}{\tt z} \ | \  \epsilon\stackrel{r}{\to}{\tt z}\in G_1\}$\\

&${\tt x}\notin fv({\tt e}_0)$&: $\{{\tt y}\stackrel{r}{\to}{\tt z}  \ |\ {\tt y}\stackrel{r}{\to}{\tt z} \in G_1\}\cup
     \{\epsilon\stackrel{\downarrow}{\to}q  \ | \ \epsilon\stackrel{r}{\to}q \in G_1\}\ \cup$ \\
&&     $\hfill 
     \{p\stackrel{\downarrow}{\to}\epsilon  \ | \ p\stackrel{r}{\to}\epsilon \in G_1\}$\\

$G^{\epsilon\mapsto{\tt x}}_2$  &stands for& 
$\{\ {\tt y}\stackrel{r}{\to}{\tt x}\  |\ \
   {\tt y}\stackrel{r}{\to}\epsilon \in G_2\}
 \cup
 \{\ \epsilon\stackrel{\downarrow}{\to}{\tt x}\  |\ \
   \epsilon\stackrel{r}{\to}\epsilon \in G_2\ \}$\\

$G\cup_{e}G'$ &stands for&the restriction of $G\cup G'$ such that 
the codomain $\subseteq fv({\tt e})\cup\{\epsilon\}$\\
\et
\bigskip

\fl First we consider how much information from $G_1$ we can preserve. We have that the whole state ${\tt e}_1{\tt@e}_2:\rho$ in left hand side for the $c$-call is strictly larger than ${\tt e}_1:\rho$. 
The variable {\tt x} is not free in $\lambda{\tt x.e}_0$ and so does not belong to the target of $G_1$. If a variable {\tt z}
$\in$   fv($\lambda{\tt x.e}_0$) is bound in 
$\rho_0$ then it is bound to the same state in $\rho_0[{\tt x}\mapsto v_2]$.
 Therefore, if there is an arc ${\tt y}\stackrel{r}{\to}{\tt z}$ in $G_1$, then it also safely describes the $c$-call and can be preserved. Also, if there is an arc $\epsilon\stackrel{r}{\to}{\tt z}$ in $G_1$, then an arc $\epsilon\stackrel{\downarrow}{\to}{\tt z}$ describes the c-call. Further, if ${\tt x}\notin fv({\tt e}_0)$ then ${\tt e}_0:\rho_0[{\tt x}\mapsto v_2]={\tt e}_0:\rho_0$ and then $\lambda{\tt x.e}_0:\rho_0\succ {\tt e}_0:\rho_0[{\tt x}\mapsto v_2]={\tt e}_0:\rho_0$. In this case, if there is an arc $p\stackrel{r}{\to}\epsilon$ going to $\epsilon$ in $G_1$, then  the arc $p\stackrel{\downarrow}{\to}\epsilon$ describes the $c$-call.\\
Now consider which information we can gain from $G_2$. We have that the whole state ${\tt e}_1{\tt@e}_2:\rho$ in left hand side for the $c$-call is strictly larger than ${\tt e}_2:\rho$.  
If ${\tt x}\in fv({\tt e}_0)$ then in ${\tt e}_0:\rho_0[{\tt x}\mapsto v_2]$ we have that ${\tt x}$ is bound to the whole state in the right hand side for the evaluation of the operand. So in this case, if there is an arc ${\tt y}\stackrel{r}{\to}\epsilon$ in $G_2$  then  the arc ${\tt y}\stackrel{r}{\to}{\tt x}$ describes the $c$-call, and if there is an arc $\epsilon\stackrel{r}{\to}\epsilon$ in $G_2$  then  the arc $\epsilon\stackrel{\downarrow}{\to}{\tt x}$ describes the $c$-call. If ${\tt x}\notin fv({\tt e}_0)$ then we cannot gain any information from $G_2$. 
The restriction built into the definition of $\cup_{e_0}$ ensures that this holds.

$$
\infer[(\textnormal{ApplyG})]
{{\tt e}_1{\tt@e}_2:\rho\Downarrow v, (G';G)}
{{\tt e}_1{\tt@e}_2:\rho\tosub{c}{\tt e}':\rho',G' &
{\tt e}':\rho'\Downarrow v,G}
$$
\smallskip

\fl The size-change graph (G';G) is the composition of the two graphs.
\edfn

\fl
In the size-change graphs generated by the rules above, the less-than relations $(x\stackrel{\downarrow}{\to}y)$ 
in (VarG)-rule arise from the sub-environment property of $\succ_1$ from 
Lemma \ref{lem-sub-env}. The remaining relations $\stackrel{\downarrow}{\to}$ arise from 
the subexpression property of $\succ_2$. The  relations based on the sub-environment property 
capture the case that the state on the right hand side is fetched 
from the environment in the left hand side. The equality relations $\stackrel{=}{\to}$ describe 
how values are preserved under calls and evaluations.

\vair
\blem\label{lem-graph-extend}

$s\to s'$ 
   {\rm(by Definition \ref{def-evaluate-and-call-relation})}
   iff 
$s\to s', G$ 
   {\rm(by Definition  \ref{def-eval-call-generate-graphs})
   for some $G$}.
Further, 
$s\Downarrow s'$
iff
$s\Downarrow s', G$   
   for some $G$.
\elem

\bthm \label{graphs_safe}
{\rm (The extracted graphs are safe)}\hfill\ \\
$s\to s', G$ or $s\Downarrow s', G$ {\rm(by Definition  \ref{def-eval-call-generate-graphs})} implies $G$ is safe for 
$(s,s')$ (with $source$ and $target$ sets extended as necessary).
\ethm

Lemma \ref{lem-graph-extend}
is immediate since the new rules extend 
the old, without any restriction on their applicability. Proof of ``safety'' 
Theorem \ref{graphs_safe} is in Appendix.\\

\vair
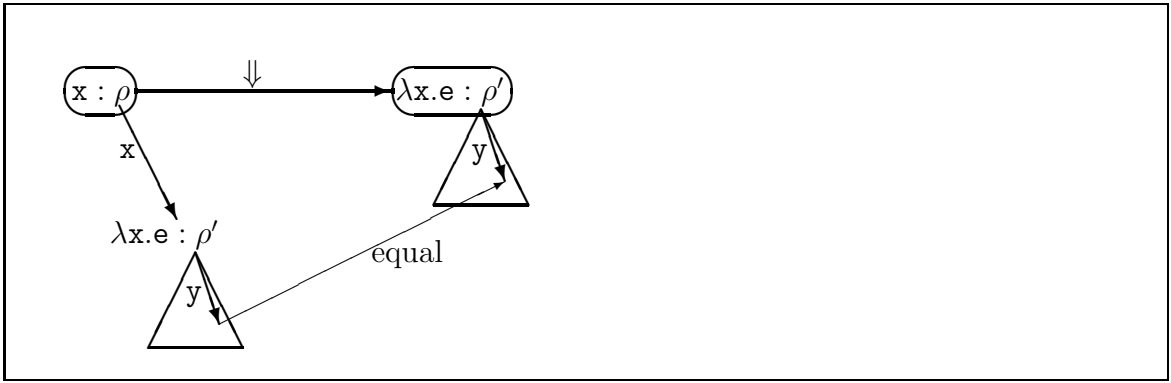
\begin{fig}{Data-flow in a variable evaluation}{h}{fig-data-flow-variable}

    \fl{\tt \large   \setlength{\unitlength}{1.8pt}

\begin{picture}(300,75)(80,-5)
\thicklines

              \put(84,52){${\tt x}:\rho$}
              \put(90,54){\oval(15,10)}
              \put(98,54){\vector(1,0){53}}
              \put(120,56){$\Downarrow$}
              \put(94,51){\vector(1,-2){12}}
              \put(94,40){x}

              \put(152,52){$\lambda{\tt x.e}:\rho'$}
              \put(164,54){\oval(25,10)}
              \put(170,50){\line(-1,-2){10}}
              \put(170,50){\line(1,-2){10}}
              \put(160,30){\line(1,0){20}}
              \put(170,50){\vector(1,-3){5}}
              \put(168,40){y}

              \put(92,22){$\lambda{\tt x.e}:\rho'$}
              \put(110,20){\line(-1,-2){10}}
              \put(110,20){\line(1,-2){10}}
              \put(100,0){\line(1,0){20}}
              \put(110,20){\vector(1,-3){5}}
              \put(108,10){y}
	      
	      \thinlines
	      
              \put(147,18){{\rm equal}}
              \put(115,5){\vector(2,1){60}}

\end{picture}}

\end{fig}

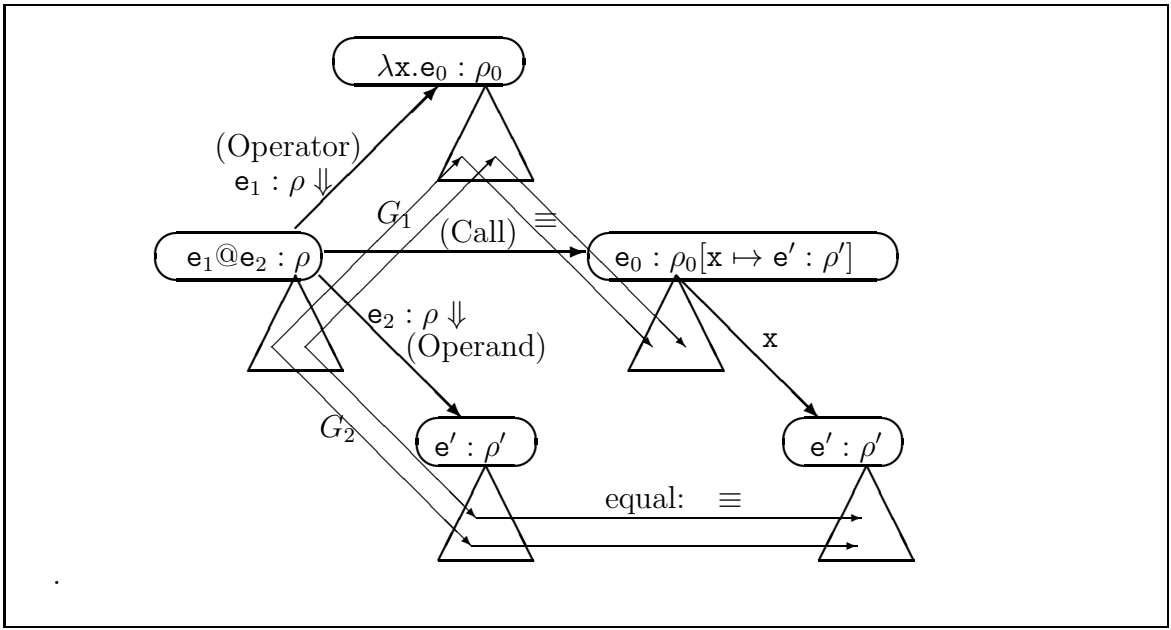
\begin{fig}{Data-flow in an application}{h}{fig-data-flow-apply}.
    \fl{\tt \large   \setlength{\unitlength}{1.8pt}
\begin{picture}(300,120)(10,25)
\thicklines

              \put(67,132){$\lambda{\tt x.e}_0:\rho_0$}
              \put(78,135){\oval(40,10)}
              \put(90,130){\line(-1,-2){10}}
              \put(90,130){\line(1,-2){10}}
              \put(80,110){\line(1,0){20}}

              \put(27,92){${\tt e}_1{\tt@e}_2:\rho$}
              \put(38,94){\oval(35,10)}
              \put(56,95){\vector(1,0){55}}
              \put(55,90){\vector(1,-1){30}}
	      
              \put(73,73){\rm (Operand)}
              \put(65,80){e$_2:\rho\Downarrow$}
              \put(50,100){\vector(1,1){30}}
	      
              \put(33,115){\rm (Operator)}
              \put(37,108){e$_1:\rho\Downarrow$}
	      
              \put(50,90){\line(-1,-2){10}}
              \put(50,90){\line(1,-2){10}}
              \put(40,70){\line(1,0){20}}
              \put(67,101){$G_1$}
	      
              \put(80,97){\rm (Call)}

              \put(117,92){${\tt e}_0:\rho_0[{\tt x}\mapsto{\tt e}':\rho']$}
              \put(144,94){\oval(65,10)}
              \put(130,90){\vector(1,-1){30}}
              \put(148,75){x}
              \put(130,90){\line(-1,-2){10}}
              \put(130,90){\line(1,-2){10}}
              \put(120,70){\line(1,0){20}}

              \put(79,52){${\tt e}':\rho'$}
              \put(88,55){\oval(25,10)}
              \put(90,50){\line(-1,-2){10}}
              \put(90,50){\line(1,-2){10}}
              \put(80,30){\line(1,0){20}}

              \put(158,52){${\tt e}':\rho'$}
              \put(165,55){\oval(25,10)}
              \put(170,50){\line(-1,-2){10}}
              \put(170,50){\line(1,-2){10}}
              \put(160,30){\line(1,0){20}}
	      
	      \thinlines
              \put(45,75){\vector(1,1){40}}
              \put(52,75){\vector(1,1){40}}
	      
              \put(45,75){\vector(1,-1){42}}
              \put(52,75){\vector(1,-1){36}}
              \put(55,56){$G_2$}
	      
              \put(85,115){\vector(1,-1){40}}
              \put(92,115){\vector(1,-1){40}}
              \put(100,100){$\equiv$}
	      
              \put(115,41){{\rm equal:} $\equiv$}
              \put(88,39){\vector(1,0){81}}
              \put(87,33){\vector(1,0){81}}

\end{picture}
}

\end{fig}

\fl The diagram of Figure \ref{fig-data-flow-variable} illustrates the data-flow in a variable evaluation.    
The diagram of Figure \ref{fig-data-flow-apply} may be of some use in 
visualising data-flow during evaluation of ${\tt e}_1{\tt @e}_2$.
States are in ovals and triangles represent environments.
In the application ${\tt e}_1  {\tt @ e}_2:\rho$ on 
the left, operator ${\tt e}_1:\rho$ evaluates to 
$\lambda {\tt x.e}_0:\rho_0,G_1$ 
and operand ${\tt e}_2:\rho$ evaluates to ${\tt e}':\rho',G_2$. 
The size-change graphs $G_1$ and $G_2$ show relations between variables 
bound in their environments. There is a call from the application 
${\tt e}_1{\tt @ e}_2:\rho$ to 
${\tt e}_0:\rho_0[{\tt x}\mapsto {\tt e}':\rho']$ 
the body of the operator-value with the environment extended with a binding of 
${\tt x}$ to the operand-value ${\tt e}':\rho'$.

It is possible to approximate the calls and evaluates to relations with different degrees of precision depending on how much information is kept about the bindings in the environment. Here we aim at a coarse approximation, where we remove all environment components.\footnote{It is possible to keep a little more information in the graphs than we do here even with no knowledge about value-bindings in the environment. 
We have chosen the given presentation for simplicity.}

\subsection{Construction of size-change graphs by abstract 
interpretation} 
\label{sec-absint-generate-graphs}

We now extend the coarse approximation to construct size-change 
graphs.

\bdfn\label{def-approx-eval-and-graph-gen} {\rm (Approximate
evaluation and call with graph generation)} \\ The judgement forms are
now ${\tt e}\to{\tt e}', G$ and ${\tt e}\Downarrow{\tt e}', G$, where
source($G$) = fv({\tt e})$\cup\{\epsilon\}$ and target($G$) = fv({\tt
e}')$\cup\{\epsilon\}$. The inference rules are:


$$
\infer[(\textnormal{ValueAG})]{\lambda{\tt x.e} \Downarrow
  \lambda{\tt x.e},{\it id}^=_{\lambda{\tt x.e}}}{}
\quad
\infer[(\textnormal{VarAG})]{{\tt x}\Downarrow v_2,\
  \{ {\tt x}\stackrel{\downarrow}{\to}{\tt y}
         \ |\ {\tt y} \in {\it fv}(v_2)\}\cup  
\{{\tt x}\stackrel{=}{\to}{\epsilon}\}}{
{\tt e}_1{\tt@e}_2\in{\it subexp}({\tt P})
  \hair
  {\tt e}_1\Downarrow\lambda{\tt x.e}_0,G_1
  \hair\hair
   {\tt e}_2\Downarrow v_2,G_2
}
$$
\smallskip

$$
\infer[(\textnormal{OperatorAG})]{{\tt e}_1{\tt@e}_2\tosub{r}{\tt e}_1,{\it id}^\downarrow_{{\tt e}_1}}
{}
\qquad
\infer[(\textnormal{OperandAG})]{{\tt e}_1{\tt@e}_2\tosub{d}{\tt e}_2,{\it id}^\downarrow_{{\tt e}_2}}
{}
$$
\smallskip

$$
\infer[(\textnormal{CallAG})]{{\tt e}_1{\tt@e}_2\tosub{c} {\tt e}_0,G_1^{-\epsilon/\lambda x.e_0}\cup_{e_0}
                                        G^{\epsilon\mapsto{\tt x}}_2}{
{\tt e}_1\Downarrow\lambda{\tt x.e}_0,G_1
  \hair\hair
  {\tt e}_2\Downarrow v_2,G_2}
\quad
\infer[(\textnormal{ApplyAG})]{{\tt e}_1{\tt@e}_2\Downarrow v,
G';G}{
{\tt e}_1{\tt@e}_2\tosub{c} {\tt e}',G'
  \hair\hair
  {\tt e}'\Downarrow v,G}
$$
\edfn
\smallskip

\blem \label{abs-graphs}Suppose ${\tt P}:[]\to^*{\tt e}:\rho$. 
If  ${\tt e}:\rho\to{\tt e}':\rho',G$ by definition \ref{def-eval-call-generate-graphs}
then ${\tt e}\to{\tt e}',G$. Further, if
${\tt e}:\rho\Downarrow{\tt e}':\rho',G$ then 
${\tt e}\Downarrow{\tt e}',G$.
\elem

\bprf  Follows from Lemma \ref{abs-super}; see the Appendix.
\eprf

\bdfn 
$$
{\it absint}({\tt P}) = \{\ G_j\ |\ j>0 \land
   \exists{\tt e}_i,G_i (0\leq i \leq j):
     {\tt P}={\tt e}_0 \land 
     ({\tt e}_0\to{\tt e}_{1},G_{1}) \land \ldots\land
     ({\tt e}_{j-1}\to{\tt e}_{j},G_{j})
     \ \}
$$

\edfn

\bthm \hfill
\be[\em(1)]
\item The set ${\it absint}({\tt P})$ is safe for {\tt P}.
\item The set  ${\it absint}({\tt P})$ can be effectively computed 
from {\tt P}.
\ee
\ethm
\bprf 
Part 1: Suppose ${\tt P}:[] = s_0\to s_1\to\ldots \to  s_j$. Theorem 
\ref{graphs_safe} implies
$s_i\to s_{i+1}, G_i$ where each $G_i$ is safe for the pair $(s_i,s_{i+1})$. 
Let $s_i = {\tt e}_i:\rho_i$. By Lemma \ref{abs-graphs},
${\tt e}_i \to {\tt e}_{i+1}, G_i$.  
By the definition of ${\it absint}({\tt P})$, $G_j\in {\it absint}({\tt P})$ .

Part 2: There is only a fixed number of subexpressions of {\tt P}, or 
of possible size-change graphs with source and target 
$\subseteq \{\epsilon\}\cup\{{\tt x}\mid{\tt x}\mbox{ is a variable in {\tt P} }\}$. 
Thus ${\it absint}({\tt P})$
can be computed by applying Definition \ref{def-approx-eval-and-graph-gen}
exhaustively, starting with {\tt P}, until no new graphs or 
subexpressions are obtained.
\eprf

\section{Some examples} \label{sec:examples}

\subsection{A simple example}

\fl Using Church numerals 
(${\tt n} = \lambda{\tt s}\lambda{\tt z.s}^n({\tt z})$), 
we expect {\tt 2 succ 0} to reduce to {\tt 
succ(succ 0)}. However this  contains unreduced redexes because 
call-by-value does not reduce under a $\lambda$, so we force the 
computation to carry on through by applying {\tt 2 succ 0} to the 
identity (twice). This gives:

\bp
2 succ 0 id1 id2 where
  succ = $\lambda$m.$\lambda$s.$\lambda$z. m s (s z)
  id1  = $\lambda$x.x
  id2  = $\lambda$y.y

\ep
After writing this out in full as a $\lambda$-expression, our 
analyser yields (syntactically sugared):\medskip

{\tt\bt{llll}
      &[$\lambda$s2.$\lambda$z2.(s2 @ (s2 @ z2))] &\hair& -- two --\\
   @ & [$\lambda$m.$\lambda$s.$\lambda$z. \fbox{15:}((m@s)@(s@z))] && -- succ 
   --\\
   @ & [$\lambda$s1.$\lambda$z1.z1]   &&  -- zero --\\
   @ & [$\lambda$x.x]                 && -- id1 --\\
   @ & [$\lambda$y.y]                 && -- id2 --
   \et}

\bp
Output of loops from an analysis of this program:

15$\to^* $15: [(m,>,m),(s,=,s),(z,=,z)], []

Size-Change Termination: Yes

\ep
The first number refers to the program point,
then comes a list of edges. 
The loop occurs because application of {\tt 2} forces the code for 
the successor function to be executed twice, with decreasing argument 
values {\tt m}. The notation for edges is a little
different from previously, here {\tt (m,>,m)} stands for
${\tt m}\stackrel{\downarrow}{\to}{\tt m}$.

\subsection{ $f n x = x + 2^n$ by Church numerals}\label{ex:fnx}

This more interesting program computes $f n x = x + 2^n$ by 
higher-order primitive 
recursion. If {\tt n} is a Church numeral then expression {\tt n g x} reduces to
{\tt g$^n($x}). Let {\tt x} be the successor function, and {\tt g} be 
a ``double application'' functional. Expressed in a readable named 
combinator form, we get: 
\bp
   f n x    where
   f n   =  if n=0 then succ else g(f(n-1)) 
   g r a =  r(ra)
\ep\smallskip

\fl As a lambda-expression (applied to values ${\tt n}=3,{\tt x} = 4)$
this can be written:\medskip

{\tt\bt{lllll}
[$\lambda$n.$\lambda$x. & n    &&& -- n --\\
   & @ & [$\lambda$r.$\lambda$a.\fbox{11:}(r@\fbox{13:}(r@a))] && -- 
   g --\\
  & @ & [$\lambda$ k.$\lambda$ s.$\lambda$ z.(s@((k@s)@z))] && - succ-\\
  & @ & x ]   && -- x --\\\\
   @ & & [$\lambda$s2.$\lambda$z2. (s2@(s2@(s2@z2))) ]     && -- 3 --\\
   @ & & [$\lambda$s1.$\lambda$z1. (s1@(s1@(s1@(s1@z1))))] && -- 4 --
   \et}
\smallskip

\fl Following is the output from program analysis. The analysis found
the following loops from a program point to itself with the associated
size-change graph and path. The first number refers to the program point,
then comes a list of edges and last a list of numbers, the other program
points that the loop passes through. 

\bp
SELF Size-Change Graphs, no repetition of graphs: 

11 $\to^*$ 11: [(r,>,r)]           []
11 $\to^*$ 11: [(a,=,a),(r,>,r)]   [13]
13 $\to^*$ 13: [(a,=,a),(r,>,r)]   [11]
13 $\to^*$ 13: [(r,>,r)]           [11,11]

Size-Change Termination: Yes
\ep

\subsection{Ackermann's function, second-order}

This can be written without recursion  using Church numerals as:
{\tt a m n} where {\tt a = $\lambda$m. m b succ} and
{\tt b =  $\lambda$g.$\lambda$n. n g (g 1)}.  
Consequently {\tt a m = b$^{\tt m}$(succ)} and
{\tt b g n = g$^{\tt n+1}$(1)}, which can be seen to agree with the 
usual first-order definition of Ackermann's function.
Following is the same as a lambda-expression applied 
to argument values {\tt m=2, n=3}, with numeric labels on some subexpressions.

\bp
($\lambda$m.m b succ) 2 3  =  ($\lambda$m.m@b@succ)@2@3
($\lambda$m.m@($\lambda$g.$\lambda$n.n@g@(g@1))@succ)@2@3
($\lambda$m.m@($\lambda$g.$\lambda$n.\fbox{9:}(n@g@\fbox{13:}(g@1)))@succ)@2@3
{\rm where}
 1   =  $\lambda$s1.$\lambda$z1. \fbox{17:}(s1@z1)
succ =  $\lambda$k.$\lambda$s.$\lambda$z. \fbox{23:}(s@\fbox{25:}(k@s@z))
 2   =  $\lambda$s2.$\lambda$z2. s2@(s2@z2)
 3   =  $\lambda$s3.$\lambda$z3. \fbox{39:}(s3@\fbox{41:}(s3@\fbox{43:}(s3@z3)))

\ep

\fl Output from an analysis of this program is shown here.\\
{\small 
(It is not always the case that the same loop is shown for all program points in its path)}

\bp
SELF Size-Change Graphs, no repetition of graphs:

 9 $\to^* $ 9: [($\epsilon$,>,n),(g,>,g)]           [13]
 9 $\to^*$  9: [(g,>,g)]                   [17]
13 $\to^*$ 13: [(g,>,g)]                   [9]
17 $\to^*$ 17: [(s1,>,s1)]                 [9]
23 $\to^*$ 23: [(k,>,k),(s,=,s),(z,=,z)]   [25]
23 $\to^*$ 23: [(s,>,s)]                   [9]
23 $\to^*$ 23: [(s,>,s),(z,>,k)]           [25,17,9]
25 $\to^*$ 25: [(k,>,k),(s,=,s),(z,=,z)]   [23]
25 $\to^*$ 25: [(s,>,s),(z,>,k)]           [17,9,23]
25 $\to^*$ 25: [(s,>,s)]                   [23,9,23]
39 $\to^*$ 39: [(s3,>,s3)]                 [9]
41 $\to^*$ 41: [(s3,>,s3)]                 [9,39]
43 $\to^*$ 43: [(s3,>,s3)]                 [9,39,41]

Size-Change Termination: Yes

\ep

\subsection{Arbitrary natural numbers as inputs}

The astute reader may have noticed a limitation in the above examples: each only concerns {\em a single $\lambda$-expression}, e.g., Ackermann's function applied to  argument values {\tt m=2, n=3}.

In an implemented version of the $\lambda$-termination analysis
a program may have an arbitrary natural number as input; this is represented by $\bullet$. 
Further, programs can have 
as constants the predecessor, successor and zero-test functions,
and if-then-else expressions. 
We show, by some examples using $\bullet$, that the size-change termination approach
can handle the Y-combinator.

\

In Section \ref{sec-regular-program-inputs} we show how to do size-change analysis 
of  $\lambda$-expressions applied to {\em sets  of  argument values} in a more classic context, using Church or other numeral notations instead of $\bullet$.

\subsection{A minimum function, with general recursion and Y-combinator}

This program computes the minimum of its two inputs using the call-by-value 
combinator 
$Y=$ {\tt$\lambda$p. [$\lambda$q.p@($\lambda$s.q@q@s)] @
                     [$\lambda$t.p@($\lambda$u.t@t@u)]}.
The program, first as a first-order recursive definition.

\bp
  m x y = if x=0 then 0 else if y=0 then 0 else succ (m (pred x) (pred y))

\ep
Now,  in $\lambda$-expression form for analysis.\medskip

{\tt\bt{lllllll}
\multicolumn{7}{l}{
\{$\lambda$p. [$\lambda$q.p@($\lambda$s.q@q@s)] @
                     [$\lambda$t.p@($\lambda$u.t@t@u)]\} \hair -- the Y combinator --}\\

@\\

[$\lambda$m.$\lambda$x.$\lambda$y.\fbox{27:}if&((ztst @ &x),\\

&0,\\

&\fbox{32:}\hair\hair if&((ztst @ y),\\

&&0,\\

&&\fbox{37:}succ @ \fbox{39:} m @ (pred@x) @ (pred@y)] \\

@ $\bullet$\\

@ $\bullet$
\et}

\bp
Output of loops from an analysis of this program:

27 $\to^*$ 27: [(x,>,x),(y,>,y)]    [32,37,39]
32 $\to^*$ 32: [(x,>,x),(y,>,y)]    [37,39,27]
37 $\to^*$ 37: [(x,>,x),(y,>,y)]    [39,27,32]
39 $\to^*$ 39: [(x,>,x),(y,>,y)]    [27,32,37]

Size-Change Termination: Yes
\ep
\vair
\subsection{Ackermann's function, second-order with constants and Y-combinator}

Ackermann's function can be written as:
{\tt a m n} where {\tt a m = b$^{\tt m}$(suc)} and
{\tt b g n = g$^{\tt n+1}$(1)}. The following program 
expresses the computations of both {\tt a} and {\tt b} by loops,
using the Y 
combinator (twice). 
\vair

{\tt\bt{llllll}
[$\lambda$ y.$\lambda$ y1. \\

(y1 @\\

$\lambda$ a.$\lambda$ m.
\fbox{11:}if( & \multicolumn{3}{l}{(ztst@m),}\\

              & \multicolumn{3}{l}{$\lambda$ v.(suc@v),}\\
	      
	      & \fbox{19:}( & (y
@\\

&&\multicolumn{3}{l}{$\lambda$ b.$\lambda$ f.$\lambda$ n.}\\

&&\fbox{25:}if( & (ztst@n),\\

&&              & \fbox{29:}(f@1),\\
		
&&              & \fbox{32:}f@\fbox{34:}b @ f @ (pred@n))\\

& \multicolumn{3}{l}{@ \fbox{41:} a @ (pred@m)}]\\\\

\multicolumn{6}{l}{
@ \{$\lambda$p.[$\lambda$q.p@($\lambda$s. q@q@s)] @
                     [$\lambda$t.p@($\lambda$u.t@t@u)]\}
}\\

\multicolumn{6}{l}{
@ \{$\lambda$p1.[$\lambda$q1.p1@($\lambda$s.\fbox{72:}q1@1q@s1)] @
                     [$\lambda$t1.p1@($\lambda$u1.\fbox{81:}t1@t1@u1)]\}
}\\

@ $\bullet$\\

@ $\bullet$
\et}

\bp
Output of loops from an analysis of this program:

SELF Size-Change Graphs no repetition of graphs: 
\ep

\bp

11 $\to^* $11: [(a,>,y),(m,>,m)]    [19,41,72]
11 $\to^* $11: [(m,>,m)]            [19,41,72,11,19,41,72]
19 $\to^* $19: [(a,>,y),(m,>,m)]    [41,72,11]
19 $\to^* $19: [(m,>,m)]            [41,72,11,19,41,72,11]
25 $\to^* $25: [(f,>,b),(f,>,f)]    [29]
25 $\to^* $25: [(f,=,f),(n,>,n)]    [32,34]
25 $\to^* $25: [(f,>,f)]            [29,25,32,34]
29 $\to^* $29: [(f,>,f)]            [25]
32 $\to^* $32: [(f,>,b),(f,>,f)]    [25]
32 $\to^* $32: [(f,=,f),(n,>,n)]    [34,25]
32 $\to^* $32: [(f,>,f)]            [25,32,34,25]
34 $\to^* $34: [(f,=,f),(n,>,n)]    [25,32]
34 $\to^* $34: [(f,>,b),(f,>,f)]    [25,29,25,32]
34 $\to^* $34: [(f,>,f)]            [25,29,25,32,34,25,32]
41 $\to^* $41: [(m,>,m)]            [72,11,19]
72 $\to^* $72: [(s1,>,s1)]          [11,19,41]
81 $\to^* $81: [(u1,>,u1)]          [11,19,41]

Size-Change Termination: Yes
\ep

\fl

\vair
\subsection{Imprecision of abstract interpretation}

It is natural to wonder 
whether the gross approximation of Definition \ref{def-eval-approx} 
comes at a cost. The (VarA) rule can in effect 
``mix up'' different function applications, losing the coordination 
between operator and operand that is present in the exact semantics.

We have observed this in practice: The first time we had programmed 
Ackermann's using explicit recursion, we used the same instance of 
Y-combinator for both 
loops, so the single Y-combinator expression was ``shared''. The analysis did 
not discover that the program terminated. 

However when this was replaced by 
the ``unshared'' version above, with two instances of the Y-combinator
({\tt y} and {\tt y1}) (one for each application), the problem disappeared 
and termination was correctly recognised.

\subsection{A counterexample to a conjecture}

Sereni disproved in \cite{sereni2,sereni} our  {\bf conjecture}  that the size-change 
method would recognise as terminating any 
simply typed $\lambda$-expression. The root of the problem is the imprecision of abstract interpretation just noted. A counter-example: the $\lambda$-expression
$$
E = (\lambda a.a(\lambda b.a(\lambda c d.d)))(\lambda e.e(\lambda f.f))
$$
is simply-typable but not size-change terminating. Its types are any
instantiation of \medskip

$\ba{lcl}
a&:&((\tau\to\tau)\to\mu\to\mu)\to\mu\to\mu\\
b,c& : &\tau\to\tau\\
d&:&\mu\\
e&:&(\tau\to\tau)\to\mu\to\mu\\
f&:&\tau
\ea$
\bigskip

\section{Arbitrary $\lambda$-regular program inputs (Extended $\lambda$-calculus)} 
\label{sec-regular-program-inputs}

Above we have analysed the termination behaviour of a single closed $\lambda$-expression. We now analyse the termination behaviour for a program in the $\lambda$-calculus for all possible inputs from a given input-set of $\lambda$-expressions (e.g., Church numerals). The first step is to define which sets of $\lambda$-expressions we consider. A well-defined input set will be the set of closed expressions in the ``language'' generated by a $\lambda$-regular grammar. 

We extend the syntax and semantics of the $\lambda$-calculus to handle expressions containing nonterminals. An extended lambda term represents all instances of a program with input taken from the input set. If our analysis certifies that the extended term terminates, then this implies that the program will terminate for all possible inputs.

\subsection{$\lambda$-regular grammars}

We are interested in a $\lambda$-regular grammar for the sake of the {\em language}  that it generates:
a set of pure $\lambda$-expressions (without nonterminals). This is done using the derivation relation $\Rightarrow^*_\Gamma$, soon to be defined.

\bdfn \label{def-lamda-reg}\hfill

\be[(1)]
\item A {\em $\lambda$-regular grammar} has form $\Gamma =(N,\Pi)$ where $N$ is a
finite set of {\em nonterminal} symbols and $\Pi$ is a finite set of {\em productions}.
\bigskip

\item A {\em $\Gamma$-extended  $\lambda$-expression} has the following syntax:\medskip

\bt{lcl}
{\tt e, P} &\ ::=\ &\  {\tt
x | A | e @ e | $\lambda$x.e}\\
{\tt A} &\ ::=\ &\  {\rm Non-terminal name, ${\tt
A}\in N$}\\
{\tt x} &\ ::=\ &\  {\rm Variable name}\\
\et\\

 \fl{\em Exp$_\Gamma$} denotes the set of $\Gamma$-extended $\lambda$-expressions. {\em Exp} denotes the set of pure $\lambda$-expressions (without nonterminals). Clearly $Exp_\Gamma\supseteq Exp$.
\bigskip

\item A production has form ${\tt A} ::= {\tt e}$ where {\tt e}
is a  $\Gamma$-extended  $\lambda$-expression. \ee
\edfn

\fl

\bdfn \label{prodstar}Let  ${\tt nt(e)}=\{{\tt X}_1,\ldots,{\tt X}_k\}$ denote the {\em multi-set} of nonterminal
occurrences in  ${\tt e}\in$ {\em Exp$_\Gamma$}. 
The {\em derivation  relation }
$ \Rightarrow^*_\Gamma\   \subseteq Exp_\Gamma\times Exp$ is  the
smallest relation such that

\be[(1)]
\item  If  ${\tt nt(e)}=\{{\tt X}_1,\ldots,{\tt X}_k\}$ and 
${\tt X}_i \Rightarrow^*_\Gamma {\tt t}_i \in Exp$ for $i=1,\ldots,k$, \\then
${\tt e} \Rightarrow^*_\Gamma  {\tt e}[{\tt t}_1/{\tt X}_1,\ldots,{\tt t}_k/{\tt X}_k]$\vair

\item If ${\tt A} ::= {\tt e} \in \Gamma$ and ${\tt e} \Rightarrow^*_\Gamma  {\tt e'}$ then ${\tt A} \Rightarrow^*_\Gamma {\tt e'}$.
\vair
\ee
\edfn

Notice that $\Rightarrow^*_\Gamma$ relates extended $\lambda$-terms to pure $\lambda$-terms.
\vair

In the above definition \ref{prodstar} ${\tt nt(e)}=\{{\tt X}_1,\ldots,{\tt X}_k\}$ denotes the multi-set of nonterminals in {\tt e} so two different ${\tt X}_i,{\tt X}_j$ may be instances of the same nonterminal {\tt A}. In the substitution ${\tt e}[{\tt t}_1/{\tt X}_1,\ldots,{\tt t}_k/{\tt X}_k]$ such two different instances of a nonterminal may be replaced by different pure $\lambda$-terms.

\begin{exa} A grammar for Church Numerals: Consider
$$\Gamma =(\{{\tt C,A}\},
\{{\tt C} ::= \lambda\,{\tt s} \lambda\,{\tt z}\,.\,{\tt A},\ 
  {\tt A} ::= {\tt z},\  {\tt A} ::= {\tt s\,@}\, {\tt A}\}
$$
Here ${\tt A} \Rightarrow^*_\Gamma {\tt v}$ iff ${\tt v}$ has form
${\tt s}^n({\tt z})$ for some $n\geq 0$. 
Clearly ${\tt C} \Rightarrow^*_\Gamma {\tt v}$ iff ${\tt v}$ has form
$\lambda\,{\tt s} \lambda\,{\tt z}\,.\,{\tt s}^n({\tt z})$ for some
$n\geq 0$. 

\end{exa}
\fl
The following  assumption makes proofs more convenient; proof is standard and so omitted.
\blem For any $\lambda$-regular grammar $\Gamma_0$ there exists an equivalent $\lambda$-regular grammar $\Gamma_1$ such that  no production in $\Gamma_1$ has form ${\tt A}::={\tt A}'$ where ${\tt A}'\in N$. We henceforth assume that all productions in a $\lambda$-regular grammar have form ${\tt A}::={\tt e}$ where 
${\tt e}\notin N$.\qed 
\elem

\bdfn \label{def-fv-subexp} In the following  ${\tt e}$ is a $\Gamma$-extended $\lambda$-expression:\hfill
\be[(1)]
\item Define the {\em free variables of {\tt e}} by $\mathit{fv}({\tt e}) = \{{\tt x}\ |\  \exists {\tt t}. {\tt e}\Rightarrow^*_\Gamma {\tt t}$ and ${\tt x}\in \mathit{fv}({\tt t})\}$ 
\vair

\item Define that ${\tt e}$ is {\em closed} iff $t$ is closed for all $t$ such that $e\Rightarrow^*_\Gamma t$.
It follows that ${\tt e}$ is closed iff $\mathit{fv}({\tt e})= \{\}$.
\vair

\item 
Define {\em subterms}({\tt e}) inductively by: \\
For a variable ${\tt x}$: $subterms({\tt x}) = \{ {\tt x}\}$.\\
For an abstraction $\lambda {\tt x}.{\tt e}$: $subterms(\lambda {\tt x}.{\tt e}) = \{\lambda {\tt x}.{\tt e}\}\cup subterms({\tt e})$.\\
For an application ${\tt e_1 @ e_2}$: $subterms({\tt e_1 @ e_2}) = \{ {\tt e_1 @  e_2}\}\cup subterms({\tt e_1})\cup subterms({\tt e_2})$.\\
For a nonterminal ${\tt A}$: $subterms({\tt A}) = \{ {\tt A}\}$.
\vair

\item Define {\em subexps}({\tt e}) as the smallest set satisfying:\\
For a variable ${\tt x}$: $subexps({\tt x}) = \{ {\tt x}\}$.\\
For an abstraction $\lambda {\tt x}.{\tt e}$: $subexps(\lambda {\tt x}.{\tt e}) = \{\lambda {\tt x}.{\tt e}\}\cup subexps({\tt e})$.\\
For an application ${\tt e_1 @ e_2}$: $subexps({\tt e_1 @ e_2}) = \{ {\tt e_1 @ e_2}\}\cup subexps({\tt e_1})\cup subexps({\tt e_2})$.\\
For a nonterminal ${\tt A}$: $subexps({\tt A}) = \{ {\tt A}\}\cup \{{\tt t}\ |\  \exists {\tt e}.{\tt A}::={\tt e} \in \Gamma$ and ${\tt t}\in subexps({\tt e})\}$.
\ee
\edfn
\fl If ${\tt e}'\in subterms({\tt e})$ then ${\tt e}'$ is syntactically present as part of ${\tt e}$.\\
If ${\tt e}'\in subexps({\tt e})$ then ${\tt e}'$ is either a subterm of ${\tt e}$ or a subexpression of a nonterminal ${\tt A}\in subterms({\tt e})$. 

Sets $subterms({\tt e}), subexps({\tt e})$ are both finite, and $subterms({\tt e})=subexps({\tt e})$ for expressions ${\tt e}$ in the pure $\lambda$-calculus.

\begin{exa} In the grammar for Church Numerals
${\tt C}$ is a closed $\Gamma$-extended expression, but
${\tt A}$ is not a closed $\Gamma$-extended expression.
Further,
$subexps({\tt A})=\{{\tt A},\ {\tt z},\ {\tt s@A},\ {\tt s}\}$, 
$subexps({\tt C})=\{{\tt C},\ \lambda\,{\tt s} \lambda\,{\tt z}\,.\,{\tt A},\ 
\lambda\,{\tt z}\,.\,{\tt A},\ {\tt A},\ {\tt z},\ {\tt s@A},\ {\tt s}\}$,
$\mathit{fv}({\tt C}) = \{\}$, $\mathit{fv}({\tt A})=\{{\tt s},{\tt z}\}$
\end{exa}

\blem \label{prod-to-form} Let ${\tt x}$ be a variable. 
If ${\tt A} \Rightarrow^*_\Gamma {\tt x}$ then  ${\tt A}::= {\tt x} \in\Gamma$.\\
If ${\tt A} \Rightarrow^*_\Gamma \lambda {\tt x}.{\tt e}$ then  there exists  ${\tt e}'\in Exp_\Gamma$ such that  ${\tt A}::= \lambda {\tt x}.{\tt e}'\in\Gamma$.\\
If ${\tt A} \Rightarrow^*_\Gamma {\tt e}_1{\tt @e}_2$ then  there exist  ${\tt e}_1',{\tt e}_2'\in Exp_\Gamma$ such that  ${\tt A}::= {\tt e}_1'{\tt @e}_2'\in\Gamma$.
\elem

Any production has one of the forms ${\tt A}::= {\tt x}$, ${\tt A}::=
\lambda {\tt x}.{\tt e}$, ${\tt A}::= {\tt e}_1{\tt @e}_2$. No
production performed on a subterm (which must be a nonterminal) can
give a new outermost syntactic term-constructor.

The following Lemma follows from the definition of free variables of
an extended expression.

\blem
For a variable ${\tt x}$: $\mathit{fv}({\tt x}) = \{ {\tt x}\}$.\\
For an abstraction $\lambda {\tt x}.{\tt e}$: $\mathit{fv}(\lambda {\tt x}.{\tt e}) = \mathit{fv}({\tt e}) \setminus \{{\tt x} \}$.\\
For an application ${\tt e}_1{\tt @e}_2$: $\mathit{fv}({\tt e}_1{\tt @e}_2) = \mathit{fv}({\tt e}_1)\cup \mathit{fv}({\tt e}_2)$.\\
For a nonterminal {\tt A} $\in$ N: $\mathit{fv}({\tt A}) = \{{\tt x}\ |\  \exists {\tt t}. {\tt A}\Rightarrow^*_\Gamma {\tt t}$ and ${\tt x}\in \mathit{fv}({\tt t})\}$. 
\elem

\blem\label{lem:fin-subexp-fv}
For ${\tt A}\in N$ the sets $subexps({\tt A})$ and $\mathit{fv}({\tt A})$ are finite and computable.
\elem

Proof is straightforward.

\subsection{Extended environment-based semantics.}

A  semantics extending Definition \ref{def-evaluate-and-call-relation-environment}
addresses the problem of substitution in expressions with non-terminals. Environments  bind $\lambda$-variables (and not  non-terminals) to values. 

\bdfn \label{def-environment-etc-ext} {\rm (Extended states, values and environments)}
{\it State}, {\it Value}, {\it Env} are the smallest
sets such that
\vspace{1mm}

\fl\bt{lccclr}

{\it State} &$\hair=\hair\{$&  ${\tt e}:\rho$
&$\hair|\hair$& ${\tt e}\in{\it Exp_\Gamma},
\rho\in{\it Env}\mbox{\ and \ }
{\mathit fv}({\tt e})\subseteq {\it dom}(\rho)$&$\}$\\

{\it Value} &$\hair=\hair\{$&  $\lambda{\tt x.e}:\rho$
&$\hair|\hair$& $
\lambda{\tt x.e}:\rho\in{\it State}$&$\}$\\

{\it Env} &$\hair=\hair\{$& $ \rho:X\to {\it Value}$
&$\hair|\hair$& $X\mbox{\  is a finite set of variables}$&$\}$
\et
\bigskip

\fl
The empty environment with  domain $X=\emptyset$ is written $[]$.
The evaluation judgement form is
$s\Downarrow v$ where $s\in{\it State}, v \in {\it Value}$.
\edfn

The following rules for calls and evaluations in the extended language are simple extensions of the rules for pure $\lambda$-calculus to also handle nonterminals.

\bdfn \label{def-environment-evaluation-ext} {\rm (Extended environment-based
evaluation)}
The judgement forms are ${\tt e }:\rho \to{\tt e}':\rho'$ and
${\tt e}:\rho \Downarrow{\tt e}':\rho'$, where ${\tt e},{\tt e'}\in Exp_\Gamma$, $e: \rho$ and $e':\rho'$ are states. 
The evaluation
and call relations $\Downarrow,\to$ are defined by the following inference rules, where
$\to\ =\tosub{r}\cup\tosub{d}\cup\tosub{c}\cup\tosub{n}$. 

$$
\infer[{\mbox{{\it \small A ::= {\tt e} $\in \Gamma$\ }}}(\textnormal{GramX})\qquad\qquad\mbox{New rule}]
{A:\rho\tosub{n} {\tt e}:\rho}
{}
$$

$$
\infer[{\mbox{{\it \small x $\in$  \{c,n\}\ }}}(\textnormal{ResultX})\qquad\qquad\mbox{Extended {\bf Def.} \ref{def-evaluate-and-call-relation-environment} (Apply)}]
{{\tt e}:\rho \Downarrow v}
{{\tt e}:\rho \tosub{x} {\tt e}':\rho'
  &
  {\tt e}':\rho'\Downarrow v}
$$


\fl The following rules have not been changed (but now expressions
belong to $Exp_\Gamma$).
\smallskip

$$
\infer[(\textnormal{ValueX})]{\lambda{\tt x.e}:\rho \Downarrow
  \lambda{\tt x.e}:\rho}{}
\qquad
\infer[{\mbox{\small $\rho({\tt x})= {\tt e}':\rho'$\ }}(\textnormal{VarX})]
{{\tt x}:\rho\Downarrow {\tt e}':\rho'}
{}
$$

$$
\infer[(\textnormal{OperatorX})]
{{\tt e}_1{\tt@e}_2:\rho \tosub{r}{\tt e}_1:\rho}
{}
\qquad
\infer[(\textnormal{OperandX})]
{{\tt e}_1{\tt@e}_2:\rho \tosub{d}{\tt e}_2:\rho}
{{\tt e}_1:\rho\Downarrow v_1}
$$

$$
\infer[(\textnormal{CallX})]
{{\tt e}_1{\tt@e}_2\tosub{c} {\tt e}_0:\rho_0[x\mapsto v_2]}
{{\tt e}_1:\rho \Downarrow\lambda{\tt x.e}_0:\rho_0
  &
  {\tt e}_2:\rho\Downarrow v_2}
$$
\edfn 

A $\Gamma$-extended {\em program} is a closed expression $ {\tt P}\in
Exp_\Gamma$.  While evaluating a program in the extended language
(${\tt P}:[]\Downarrow \_$), all calls and subevaluations will be from
state to state.

In pure $\lambda$-calculus the evaluation relation is
deterministic. The extended language is nondeterministic since a
nonterminal ${\tt A}$ may have ${\tt A}::={\tt e}$ for more than one
${\tt e}$.

Informally explained, consider closed extended $\lambda$-expression
${\tt e}{\tt @B}$ where nonterminal ${\tt B}$ satisfies
$\mathit{fv}({\tt B})=\{\}$. Then ${\tt e}{\tt @B}$ represents
application of ${\tt e}$ to all possible inputs generated by {\tt
B}. The analysis developed below can safely determine that ${\tt e}$
terminates on all inputs by analysing ${\tt e}{\tt @B}$.

 If a program in the extended language takes more than one input at a time, 
then we may rename the nonterminals and bound variables similarly as in $\alpha$-conversion. 
As an example, if a program takes two Church numerals as input, then they can be given by two grammars identical in structure:

\bt{lll}
${\tt C}_1::= \lambda {\tt s}_1.\lambda {\tt z}_1.{\tt A}_1$&${\tt A}_1::= {\tt z}_1$&${\tt A}_1::= {\tt s}_1{\tt @A}_1$ and\\
${\tt C}_2::= \lambda {\tt s}_2.\lambda {\tt z}_2.{\tt A}_2$&${\tt A}_2::= {\tt z}_2$&${\tt A}_2::= {\tt s}_2{\tt @A}_2$\\
\et

\fl and we can analyse the termination behaviour for $({\tt e}{\tt @C}_1){\tt @C}_2$. 
Such renaming can sometimes make the termination analysis more precise.

\bdfn Suppose ${\tt e}$ is a closed $\Gamma$-extended expression and ${\it nt}({\tt e})=\{{\tt A}_1,\ldots,{\tt A}_k\}$ where
$\Gamma =(N,\Pi)$ is a  $\lambda$-regular grammar.
By definition {\tt e} is {\em $\Gamma$-terminating} iff
$$
{\tt e}[{\tt t}_1/{\tt A}_1,\ldots,{\tt t}_k/{\tt A}_k]:[] \Downarrow
$$
for all pure $\lambda$-expressions ${\tt t}_1,\ldots,{\tt t}_k$ such that
${\tt A}_i\Rightarrow^*_\Gamma {\tt t}_i$ for
$i=1,\ldots,k$.
\edfn
\vair

The following rules for calls and evaluations with size-change graphs in the extended language are simple extensions of the rules for pure $\lambda$-calculus to also handle nonterminals. 

\bdfn \label{def-environment-evaluation-extb} {\rm (Environment-based
evaluation and call semantics utilizing size-change graphs)}
The judgement forms are ${\tt e }:\rho \to{\tt e}':\rho', G$ and
${\tt e}:\rho \Downarrow{\tt e}':\rho', G$, where ${\tt e},{\tt e'}\in Exp_\Gamma$, $e: \rho$ and $e':\rho'$ are states, $source(G) = \mathit{fv}(e)\cup \{\epsilon\}$ and  $target(G)=\mathit{fv}(e')\cup \{\epsilon\}$.
The evaluation
and call relations $\Downarrow,\to$ are defined by the following inference rules, where
$\to\ =\tosub{r}\cup\tosub{d}\cup\tosub{c}\cup\tosub{n}$. 

$$
\infer[{\mbox{{\it \small A ::= {\tt e} $\in \Gamma$\ }}}(\textnormal{GramG})\qquad\qquad\mbox{New rule}]
{A:\rho\tosub{n} {\tt e}:\rho, id_e^=}
{}
$$

$$
\infer[{\mbox{{\it \small x $\in$  \{c,n\}\ }}}(\textnormal{ResultG})\qquad\qquad\mbox{Extended {\bf Def.} \ref{def-eval-call-generate-graphs} (ApplyG)}]
{{\tt e}:\rho \Downarrow v,G';G}
{{\tt e}:\rho \tosub{x} {\tt e}':\rho',G'
  &
  {\tt e}':\rho'\Downarrow v,G}
$$

\fl The following rules have not been changed (but now expressions belong to $Exp_\Gamma$).

$$
\infer[(\textnormal{ValueG})]{\lambda{\tt x.e}:\rho \Downarrow
  \lambda{\tt x.e}:\rho,{\it id}^=_{\lambda{\tt x.e}}}{}
$$

$$
\infer[{\mbox{\small $\rho({\tt x})= {\tt e}':\rho'$\ }}(\textnormal{VarG})]
{{\tt x}:\rho\Downarrow e':\rho',\{{\tt x}\stackrel{=}{\to}{\epsilon}\}
  \cup\{ {\tt x}\stackrel{\downarrow}{\to}{\tt y}
         \ |\ {\tt y} \in {\mathit fv}(e')\}}
{}
$$

$$
\infer[(\textnormal{OperatorG})]
{{\tt e}_1{\tt@e}_2:\rho \tosub{r}{\tt e}_1:\rho,{\it id}^\downarrow_{{\tt e}_1}}
{}
\qquad
\infer[(\textnormal{OperandG})]
{{\tt e}_1{\tt@e}_2:\rho \tosub{d}{\tt e}_2:\rho,{\it id}^\downarrow_{{\tt e}_2}}
{{\tt e}_1:\rho\Downarrow v_1}
$$

$$
\infer[(\textnormal{CallG})]
{{\tt e}_1{\tt@e}_2\tosub{c} {\tt e}_0:\rho_0[x\mapsto v_2],G_1^{-\epsilon/\lambda x.e_0}\cup_{e_0} G^{\epsilon\mapsto{\tt x}}_2}
{{\tt e}_1:\rho \Downarrow\lambda{\tt x.e}_0:\rho_0,G_1
  &
  {\tt e}_2:\rho\Downarrow v_2,G_2}
$$
\edfn 
\medskip

\bthm \label{graphs_safeG_Ext}
{\rm (The extracted graphs are safe)}\hfill
$s\to s', G$ or $s\Downarrow s', G$ implies $G$ is safe for 
$(s,s')$. 
\ethm

\bprf 
\fl
This is shown by a case analysis as in the pure $\lambda$-calculus. For the (GramG) rule it is immediate from the definition of free variables for non-terminals.
\eprf

\subsection{Relating extended and pure $\lambda$-calculus}\ 

\fl The aim is now to show that execution of a program ${\tt P}$ in the extended language can 
simulate execution of any program ${\tt Q}$ in the pure $\lambda$-calculus, where ${\tt Q}$ is derived from ${\tt P}$ by replacing each nonterminal occurrence ${\tt A}$ in ${\tt P}$ with a pure $\lambda$-expression ${\tt A}$ can produce. The converse does not hold: it is possible that there are simulated executions that do not correspond to any instantiated program {\tt Q}. We have however  certified a number of programs to terminate when applied to arbitrary Church numerals. An example is given at the end of this section.
\smallskip

\fl \textbf{Properties of the relation $\Rightarrow^*_\Gamma$}\\
$\Rightarrow^*_\Gamma$ relates expressions ${\tt e}'\in Exp_\Gamma$ in
the extended language to expressions ${\tt e}\in Exp$ in the pure
lambda-calculus.  Notice that there are only the following possible
forms of $\Rightarrow^*_\Gamma$-related expressions:  

\bt{rrrrr}
${\tt x} \Rightarrow^*_\Gamma {\tt x}$ & &
$\lambda {\tt x}.{\tt e}'\Rightarrow^*_\Gamma \lambda {\tt x}.{\tt e}$ & &
${\tt e}_1'{\tt @e}_2'\Rightarrow^*_\Gamma {\tt e}_1{\tt @e}_2$\\
${\tt A}\Rightarrow^*_\Gamma {\tt x}$ & &
${\tt A}\Rightarrow^*_\Gamma \lambda {\tt x}.{\tt e}$ & &
${\tt A}\Rightarrow^*_\Gamma {\tt e}_1{\tt @e}_2$\\
\et
\medskip

\fl The relation $\Rightarrow^*_\Gamma$ has the following inductive properties:\medskip

\fl${\tt A}\Rightarrow^*_\Gamma {\tt t}$, for ${\tt A}\in N$ is given by definition \ref{prodstar}.\\
${\tt x}\Rightarrow^*_\Gamma {\tt x}$, -- a variable ${\tt x}$ corresponds to the same variable ${\tt x}$ and nothing else. \\
$\lambda {\tt x}.{\tt e}'\Rightarrow^*_\Gamma\lambda {\tt x}.{\tt e}$, iff ${\tt e}'\Rightarrow^*_\Gamma {\tt e}$, same ${\tt x}$.\\
${\tt e}_1'{\tt @e}_2'\Rightarrow^*_\Gamma {\tt e}_1{\tt @e}_2$ iff ${\tt e}_1'\Rightarrow^*_\Gamma {\tt e}_1$ and ${\tt e}_2'\Rightarrow^*_\Gamma {\tt e}_2$.\medskip

\blem If ${\tt e}'\Rightarrow^*_\Gamma {\tt e}$ then $\mathit{fv}({\tt e}')\supseteq \mathit{fv}({\tt e})$.
\elem

\bprf This is by induction on the structure of ${\tt e}'$.\\
Case ${\tt x}\Rightarrow^*_\Gamma {\tt x}$, immediate.\\ 
Case ${\tt A}\Rightarrow^*_\Gamma {\tt t}$ where ${\tt A}\in N$. By definition $\mathit{fv}({\tt A}) = \{{\tt x}|\exists {\tt t}. {\tt A}\Rightarrow^*_\Gamma {\tt t}$ and ${\tt x}\in \mathit{fv}({\tt t})\}$.\\
Case $\lambda {\tt x}.{\tt e}'\Rightarrow^*_\Gamma \lambda {\tt x}.{\tt e}$, iff ${\tt e}'\Rightarrow^*_\Gamma {\tt e}$. By induction the lemma holds for ${\tt e}'$ and ${\tt e}$.
Therefore $\mathit{fv}(\lambda {\tt x}.{\tt e}')= \mathit{fv}({\tt e}')\setminus\{{\tt x}\}\supseteq \mathit{fv}({\tt e})\setminus\{{\tt x}\}= \mathit{fv}(\lambda {\tt x}.{\tt e})$.\\ 
Case ${\tt e}_1'{\tt @e}_2'\Rightarrow^*_\Gamma {\tt e}_1{\tt @e}_2$, iff ${\tt e}_1'\Rightarrow^*_\Gamma {\tt e}_1$ and ${\tt e}_2'\Rightarrow^*_\Gamma {\tt e}_2$. By induction the lemma holds for ${\tt e}_1',{\tt e}_1$ and ${\tt e}_2',{\tt e}_2$.
Hence $\mathit{fv}({\tt e}_1'{\tt @e}_2')= \mathit{fv}({\tt e}_1') \cup \mathit{fv}({\tt e}_2')\supseteq \mathit{fv}({\tt e}_1) \cup \mathit{fv}({\tt e}_2)= \mathit{fv}({\tt e}_1{\tt @e}_2)$.
\eprf

If ${\tt e}\in Exp$, i.e., no nonterminals occur in ${\tt e}$, then
${\tt e}\Rightarrow^*_\Gamma {\tt e}$.  If ${\tt
A}\Rightarrow^*_\Gamma {\tt e}$ then there exist ${\tt t}\notin N$
such that ${\tt A}::={\tt t}$ and ${\tt t}\Rightarrow^*_\Gamma {\tt
e}$.

\bdfn \textbf{The relation $S$ between states}\\
Define the relation $S$ between states in the extended language and states in the pure $\lambda$-calculus as the smallest relation $S$ such that: 
\bi
\item[]$S({\tt e}':\rho',{\tt e}:\rho)$ if ${\tt e}'\Rightarrow^*_\Gamma {\tt e}$ and for all ${\tt x}\in \mathit{fv}({\tt e})$ it holds that $S(\rho'({\tt x}),\rho({\tt x}))$.
\ei
\edfn

If ${\tt e}:\rho$ is a state in the pure lambda calculus then it is also a state in the extended language and $S({\tt e}:\rho,{\tt e}:\rho)$.

\blem \label{push-S}
If $S({\tt A}:\rho',{\tt e}:\rho)$ and ${\tt A}::= {\tt t},\ {\tt t}\Rightarrow_\Gamma^* {\tt e} $ then also $S({\tt t}:\rho',{\tt e}:\rho)$.\qed
\elem

We now define a relation $T$ between size-change graphs. The intention is that $T(G',G)$ is to hold when the only difference in the generation of the graphs is due to nonterminals that take the place of pure lambda expressions.
\bdfn \textbf{The relation $T$ between size-change graphs}

Define $T(G',G)$ to hold iff 
\be[i)]
\item $source(G')\supseteq source(G)$ and $target(G')\supseteq target(G)$.
\item The subgraph of $G'$ restricted to $source(G)$ and $target(G)$ is a subset of $G$.
\item Furthermore if ${\tt z}\in source(G')\setminus source(G)$ then either there is no edge from ${\tt z}$ in $G'$ or the only edge from ${\tt z}$ in $G'$ is 
$({\tt z}\stackrel{=}{\to}{\tt z})$, and if $({\tt z}\stackrel{=}{\to}{\tt z})\in G'$ then ${\tt z}\notin target(G)$. 
\ee
\edfn

\noindent We have that $T(G_0',G_0)$, $T(G_1',G_1)$, target($G_0'$) = source($G_1'$) and target($G_0$) = source($G_1$) together imply  that $ T((G_0';G_1'),(G_0;G_1))$ holds.

\blem \label{simulation}\rm{\bf Simulation Property}
\be[i)]
\item If $S({\tt e}':\rho',{\tt e}:\rho)$ and ${\tt e}:\rho\Downarrow {\tt e}_0:\rho_0,G$ then there exist ${\tt e}_0':\rho_0',G'$ with $S({\tt e}_0':\rho_0',{\tt e}_0:\rho_0)$ and $T(G',G)$ such that ${\tt e}':\rho'\Downarrow {\tt e}_0':\rho_0',G'$.
\item If $S({\tt e}':\rho',{\tt e}:\rho)$ and ${\tt e}:\rho\tosub{x} {\tt e}_0:\rho_0,G$ with $x\in\{r,d,c\}$ then there exist ${\tt e}_0':\rho_0',G'$ and possibly $s$ such that 
either ${\tt e}':\rho'\tosub{x} {\tt e}_0':\rho_0',G'$ or ${\tt e}':\rho'\tosub{n} s\tosub{x} {\tt e}_0':\rho_0',G'$ with $S({\tt e}_0':\rho_0',{\tt e}_0:\rho_0)$, $T(G',G)$, and in the last case $S(s,{\tt e}:\rho)$.\\ 
The composite size-change graph for the double-call $ {\tt e}':\rho'\tosub{n}\  s\tosub{x} {\tt e}_0':\rho_0'$ will have the same edges as $G'$ because the $\tosub{n}$ call generates an $id^=$ graph.
\ee
\elem

\bcor For programs ${\tt P}\in Exp_\Gamma$ and ${\tt Q}\in Exp$ with ${\tt P}\Rightarrow^*_\Gamma {\tt Q}$ it holds that:\\ 
If ${\tt Q}:[]\to^* {\tt e}:\rho$ then there exists ${\tt e}':\rho'$ such that ${\tt P}:[]\to^*{\tt e}':\rho'$ and $S({\tt e}':\rho',{\tt e}:\rho)$.\\
If ${\tt Q}:[]\Downarrow {\tt e}:\rho$ then there exist ${\tt e}':\rho'$ such that ${\tt P}:[]\Downarrow {\tt e}':\rho'$ and $S({\tt e}':\rho',{\tt e}:\rho)$.
\ecor

Also notice that if ${\tt e}_1:\rho_1\tosub{n} {\tt e}_2:\rho_2$ then $\mathit{fv}({\tt e}_1)\supseteq \mathit{fv}({\tt e}_2)$ by the definition of free variables for nonterminals. (By definition, $S({\tt e}':\rho',{\tt e}:\rho)$ implies $\mathit{fv}({\tt e}')\supseteq \mathit{fv}({\tt e})$.)

\bprf Lemma \ref{simulation} is shown by induction on the tree for the proof of evaluation or call in the pure $\lambda$-calculus and uses the observation about free variables. Proof is in the appendix.
\eprf

\subsection{The subexpression property}

\bdfn Given a state $s$ in the extended language, we define its
{\em expression support} ${\it exp\_sup}(s)$ by
\[{\it exp\_sup}({\tt e}:\rho) = {\it subexps}({\tt e}) \cup
  \bigcup_{{\tt x}\in{\mathit fv}({\tt e})}
     {\it exp\_sup}(\rho({\tt x}))
     \]
\edfn

\blem \label{lem-subexpression-ext}
{\rm (Subexpression property)} If $s\Downarrow s'$ or $s\to s'$ then
${\it exp\_sup}(s)\supseteq {\it exp\_sup}(s')$.
\elem
\bcor If
${\tt P}:[]\Downarrow \lambda{\tt x.e}:\rho$ then
$\lambda{\tt x.e}\in{\it subexp}({\tt P})$. If
${\tt P}:[]\to^* {\tt e}:\rho$ then
${\tt e}\in{\it subexps}({\tt P})$.
\ecor

The proof of Lemma \ref{lem-subexpression-ext} follows the same lines
as the proof of Lemma \ref{lem-subexpression}. The proof for the rule
(Gram) is immediate from the definition of subexpressions in the
extended language. Proof omitted.

\subsection{Approximate extended semantics with size-change graphs}
\bdfn
\label{def-Ext-eval-approx}
{\rm(Approximate evaluation and call rules for extended semantics with size-change graphs)}. 
The judgement forms are now  ${\tt e}\to{\tt e}', G$ and
${\tt e}\Downarrow{\tt e}', G$, where ${\tt e},{\tt e}'\in Exp_\Gamma$, and $source(G) = \mathit{fv}(e)\cup \{\epsilon\}$ and  $target(G)=\mathit{fv}(e')\cup \{\epsilon\}$.

$$
\infer[{\mbox{{\it \small A ::= {\tt e} $\in \Gamma$\ }}}(\textnormal{GramAG})]
{A\tosub{n} {\tt e}, id_e^=}
{}
\qquad
\infer[{\mbox{{\it \small x $\in$  \{c,n\}\ }}}(\textnormal{ResultAG})]
{{\tt e}\Downarrow v,
G';G}
{{\tt e}\tosub{x} {\tt e}',G'
  &
  {\tt e}'\Downarrow v,G}
$$

$$
\infer[(\textnormal{VarAG})]
{{\tt x}\Downarrow v_2,
  \{{\tt x}\stackrel{=}{\to}{\epsilon}\}
  \cup\{ {\tt x}\stackrel{\downarrow}{\to}{\tt y}
         \ |\ {\tt y} \in {\mathit fv}(v_2)\}}
{{\tt e}_1{\tt@e}_2\in{\it subexps}({\tt P})
  &
  {\tt e}_1\Downarrow\lambda{\tt x.e}_0,G_1
  &
   {\tt e}_2\Downarrow v_2,G_2}
$$

$$
\infer[(\textnormal{OperatorAG})]
{{\tt e}_1{\tt@e}_2\tosub{r}{\tt e}_1,{\it id}^\downarrow_{{\tt e}_1}}
{}
\qquad
\infer[(\textnormal{OperandAG})]
{{\tt e}_1{\tt@e}_2\tosub{d}{\tt e}_2,{\it id}^\downarrow_{{\tt e}_2}}
{}
$$

$$
\infer[(\textnormal{ValueAG})]{\lambda{\tt x.e} \Downarrow
  \lambda{\tt x.e},{\it id}^=_{\lambda{\tt x.e}}}{}
\qquad
\infer[(\textnormal{CallAG})]
{{\tt e}_1{\tt@e}_2\tosub{c} {\tt e}_0,G_1^{-\epsilon/\lambda x.e_0}\cup_{e_0}
                                        G^{\epsilon\mapsto{\tt x}}_2}
{{\tt e}_1\Downarrow\lambda{\tt x.e}_0,G_1
  &
  {\tt e}_2\Downarrow v_2,G_2}
$$
\edfn
\medskip

Putting the pieces together, we now show how to analyse any program in the
regular grammar-extended $\lambda$-calculus . Let ${\tt P}$ be a program in the extended language.
\bdfn\ 

\fl ${\it absintExt}({\tt P}) = $
$$\{\ G_j\ |\ j>0 \land
   \exists{\tt e}_i,G_i,(0\leq i \leq j):
     {\tt P}={\tt e}_0 \land 
     ({\tt e}_0\to{\tt e}_{1},G_{1}) \land \ldots\land
     ({\tt e}_{j-1}\to{\tt e}_{j},G_{j})
     \ \}
$$

\edfn

\bthm The set  ${\it absintExt}({\tt P})$ can be effectively computed 
from {\tt P}.\ethm
\bprf 
In the extended $\lambda$-calculus there is only a fixed number of subexpressions of {\tt P}, and
a fixed number of  
of possible size-change graphs with 
$${\rm source, target} \subseteq \{\epsilon\}\cup\{{\tt x}\ |\ {\tt x} \mbox{\  is a variable that occurs in a subexpression of ${\tt P}$}\}$$
Thus ${\it absintExt}({\tt P})$
can be computed in finite time by applying Definition \ref{def-Ext-eval-approx}
exhaustively, starting with {\tt P}, until no new graphs or 
subexpressions are obtained.
\eprf

\subsection{Simulation properties of approximate extended semantics}

We will show the following properties of approximate extended semantics: 
\be[(1)]
\item Calls and evaluations for a program in {\em extended semantics with environments} can be stepwise simulated by {\em approximate extended semantics} with identical size-change graphs associated with corresponding calls and evaluations. To a call or evaluation in the extended $\lambda$-calculus with environments corresponds the same call or evaluation with environments removed.
\item 
Suppose  ${\tt P}\Rightarrow^*_\Gamma {\tt Q}$ for programs  {\tt P},{\tt Q}. Then calls and evaluations for ${\tt Q}$ in the pure lambda calculus with environments can be simulated by calls and evaluations in the approximate extended semantics for ${\tt P}$ using the relations  $\Rightarrow^*_\Gamma $ 
and $T$.
\item The extra edges in the size-change graphs in extended semantics can never give rise to incorrect termination analysis.
\ee

\blem \label{embExt}
Let ${\tt P}$ be a program in the extended language and ${\tt P}:[]\to^*{\tt e}:\rho$.\\ 
If ${\tt e}:\rho\to {\tt e}_0:\rho_0,G$ then ${\tt e}\to {\tt e}_0,G$ in approximate semantics.\\
If ${\tt e}:\rho\Downarrow {\tt e}_0:\rho_0,G$ then ${\tt e}\Downarrow {\tt e}_0,G$ in approximate semantics.
\elem
\bprf
The proof is similar to the proof for approximation of the pure lambda-calculus \ref{abs-super} and \ref{abs-graphs}.
For rules (Value), (Operator), (Operand) it is immediate. The (Gram)-rule do not refer to the environment, hence the lemma holds if the (Gram)-rule has been applied. For rules (Call) and (Result) it holds by induction. For the (Var)-rule we need induction on the total size of the derivation, and we can argue as in the case of the pure lambda calculus.
\eprf

\blem \label{simulation-approx}
Let ${\tt P}$ be a program in the extended language and ${\tt Q}$ a program in the pure $\lambda$-calculus with ${\tt P}\Rightarrow^*_\Gamma {\tt Q}$.

If ${\tt Q}:[]\to^*{\tt e}:\rho$ and ${\tt e}:\rho\Downarrow {\tt e}_0:\rho_0,G$ then there 
exist ${\tt e}',{\tt e}_0',G'$ with ${\tt e}'\Rightarrow^*_\Gamma {\tt e}$, ${\tt e}_0'\Rightarrow^*_\Gamma {\tt e}_0$, $T(G',G)$ such that ${\tt P}\to^*{\tt e}'$ and ${\tt e}'\Downarrow {\tt e}_0',G'$.

If ${\tt Q}:[]\to^*{\tt e}:\rho$ and ${\tt e}:\rho\tosub{x} {\tt e}_0:\rho_0,G,\ x\in\{r,d,c\}$ then there 
exist ${\tt e}',{\tt e}_0',G'$ with ${\tt e}'\Rightarrow^*_\Gamma {\tt e}$ ,${\tt e}_0'\Rightarrow^*_\Gamma {\tt e}_0$, $T(G',G)$ such that ${\tt P}\to^*{\tt e}'$ and either ${\tt e}'\tosub{x} {\tt e}_0',G'$ or ${\tt e}'\tosub{n}{\tt e}''\tosub{x} {\tt e}_0',G'$ where in the last case $G'$ is the composite size-change graph for the double call.
\elem

\bprf The lemma follows from the simulation property lemma \ref{simulation} together with lemma \ref{embExt}.
\eprf

\bthm\label{gamma-termination}\ 
\be[\em(1)]
\item Let ${\tt P}$ be a program in the extended language. If there is a program ${\tt Q}$ in the pure lambda-calculus such that  ${\tt P}\Rightarrow^*_\Gamma {\tt Q}$ and there exists an infinite call-sequence 
in the call-graph for ${\tt Q}$ in the exact semantics, then there exists an infinite call-sequence with no infinitely descending thread in the call-graph for ${\tt P}$ in the approximate extended semantics.
\item It follows that if each infinite call-sequence  in the call-graphs for ${\tt P}$ in the approximate extended semantics has an infinitely descending thread, then ${\tt P}$ is $\Gamma$-terminating.
\end{enumerate}
\ethm

\bprf 
(1): Assume an infinite call-sequence 
exists in the call-graph for ${\tt Q}$. By the safety of the size-change graphs in the pure $\lambda$-calculus, the size-change graphs associated with this call sequence cannot have an infinitely descending thread. By lemma \ref{simulation-approx} there exists a simulating call-sequence in the call-graph for ${\tt P}$ such that  the corresponding size-change graphs are in the $T$-relation. 
Let $G_P,G_Q $ be any such two corresponding $T$-related size-change graphs from these call-sequences, $T(G_P,G_Q)$. 
By the definition of the $T$-relation it holds that the largest subgraph of $G_P$, with $source$ and $target$ the same as  $source(G_Q)$ and $target(G_Q)$,  is equal to or a subset of $G_Q$. 
 We need to show that the possible extra variables in the size-change graphs for the simulating sequence in the call-graph for ${\tt P}$ can never take part in an infinitely descending thread. By the definition of the $T$-relation it holds that 
an edge leaving from such a variable ${\tt x}$ must have have the form $({\tt x}\stackrel{=}{\to}{\tt x})$ if any exists in the simulating sequence. Also by the definition of the $T$-relation, if $T(G_P,G_Q)$ and $({\tt x}\stackrel{=}{\to}{\tt x})\in G_P$ then ${\tt x}\notin codomain(G_Q)$. Hence either an extra thread in the size-change graphs going out from ${\tt x}$ will be finite or it will be infinitely equal ${\tt x}\stackrel{=}{\to}{\tt x}\stackrel{=}{\to}{\tt x}\stackrel{=}{\to}\ldots$, i.e. an extra variable can never take part in an infinitely descending thread in the simulating sequence.\\
(2) is a corollary to (1).
\eprf

\begin{exa}
The following is an example of a  program certified to terminate  by our proof method. The program computes $x + 2^n$ when applied to two arbitrary Church numerals for $x$ and $n$. In Section \ref{sec:examples} we analysed the program applied to Church numerals 3 and 4 (Example \ref{ex:fnx}).
\medskip

\fl Grammar for Church numerals:\hair\hair 
C ::= $\lambda {\tt s}.\lambda {\tt z}.$A\hair\hair
A ::= {\tt z}  $|$ {\tt s}{\tt @}A
\vair

\fl The program applied to two Church numerals:
\vair

{\tt\bt{lllll}
[$\lambda$n$_1$.$\lambda$n$_2$. & n$_1$    &&& -- n --\\
   & @ & [$\lambda$r.$\lambda$a.\fbox{11:}(r@\fbox{13:}(r@a))] && -- 
   g --\\
  & @ & [$\lambda$ k.$\lambda$ p.$\lambda$ q.(p@((k@p)@q))] && - succ-\\
  & @ & n$_2$ ]   && -- x --\\\\
   @ & & C  && -- Church numeral --\\
   @ & & C  && -- Church numeral --
   \et}
\bigskip

Following is the output from program analysis. The analysis found
the following loops from a program point to itself with the associated
size-change graph and path. The first number refers to the program point,
then comes a list of edges and last a list of numbers, the other program
points that the loop passes through. The program points are found automatically by the analysis. The program points 30 and 32 are not written into the presentation of the program because they involve the subexpression A of a Church numeral. The subexpression associated with 30 is A and the subexpression associated with 32 is {\tt s@}A. The loops  from 30 to itself and from 32 to itself in the output correspond to the call sequence 
A$\to${\tt s@}A$\to$A$\to${\tt s@}A\ldots.

\bp
SELF SCGS no repetition of graphs: 

11 $\to^*$ 11: [(r,>,r)]                                     []
11 $\to^*$ 11: [(a,=,a),(r,>,r)]                             [13]
13 $\to^*$ 13: [(a,=,a),(r,>,r)]                             [11]
13 $\to^*$ 13: [(r,>,r)]                                     [11,11]
30 $\to^*$ 30: [($\epsilon$,>,$\epsilon$),(s,=,s),(z,=,z)]   [32]
32 $\to^*$ 32: [($\epsilon$,>,$\epsilon$),(s,=,s),(z,=,z)]   [30]

Size-Change Termination: Yes
\ep

\

\end{exa}
\section{Concluding matters}

We have developed a method based on The Size-Change Principle to show termination of a closed expression in the untyped $\lambda$-calculus. This is further developed to analyse if a program in the $\lambda$-calculus will terminate when applied to any input from a given input set defined by a tree grammar. The analysis is safe and the method can be  completely automated. We have a simple first implementation. The method certifies termination of many interesting recursive programs, including programs with mutual recursion and parameter exchange.

\subsubsection*{Acknowledgements.} The authors gratefully acknowledge 
detailed and constructive comments by Arne Glenstrup, Chin Soon 
Lee and Damien Sereni, and insightful comments by Luke Ong, David Wahlstedt and Andreas Abel.
\appendix

\section{Proof of Lemma \ref{lem-NIS-standard}}

\bprf 
$\Rightarrow$: Assume ${\tt P}\Downarrow$. To show: CT has no infinite call chain starting with ${\tt P}$. 
The proof is by induction on the height of the proof tree. Each call rule of  
\ref{lem-NIS-standard} is associated 
with a use of rule (ApplyS) from Definition 
\ref{def-call-by-value-evaluation}. So if ${\tt P}$ is a value, there is no call from ${\tt P}$. If ${\tt P}\Downarrow$ is concluded by rule (ApplyS), then ${\tt P}={\tt e}_1{\tt@e}_2$ and  by induction there is no infinite call chain starting with ${\tt e}_1$, ${\tt e}_2$ and ${\tt e}_0[v_2/{\tt x}]$. All call chains starting with ${\tt P}$ go  directly to one of these. So, there are no infinite call chains starting with ${\tt P}$.

 $\Leftarrow$: Assume CT has no infinite call chain starting with ${\tt P}$. To show: ${\tt P}\Downarrow$.  
Since the call tree is finitely branching,
 by K\"{o}nig's lemma the whole call tree is finite, and 
 hence there exists a finite number $m$ bounding the length of all branches.

We prove  that  ${\tt e} \Downarrow$ for any expression in the call tree, by induction on the maximal length $n$ of a call chain from 
${\tt e}$. 

$n=0:$ {\tt e} is an abstraction that evaluates to itself.

$n>0:$ {\tt e} must be an application ${\tt e}={\tt e}_1{\tt@e}_2$.
By rule (Operator) 
there is a call 
${\tt e}_1{\tt@e}_2 \tosub{d} {\tt e}_1$, 
and the maximal length of 
a call chain from ${\tt e}_1$ is less than $ n$. 
By induction there exists 
$v_1$ such that ${\tt e}_1\Downarrow v_1$. 
We  now  conclude by rule (Operand) 
that
${\tt e}_1{\tt@e}_2 \tosub{r} {\tt e}_2$. By induction
there exists $v_2$ such that ${\tt e}_2\Downarrow v_2$. 

All values are abstractions, so we can write 
$v_1 =\lambda {\tt x.e}_0$.
We  now conclude by rule (Call) 
that
${\tt e}_1{\tt@e}_2 \tosub{c} {\tt e}_0[v_2/{\tt x}]$. 
 By induction again, $ {\tt e}_0[v_2/{\tt x}]\Downarrow v$
 for some $v$.
This gives us all premises for the (ApplyS) rule of
Definition \ref{def-call-by-value-evaluation}, so
 ${\tt e}= {\tt e}_1{\tt@e}_2\Downarrow v$.
\eprf

\section{Proof of Lemma \ref{abs-super}}

\bprf  To be shown:
\hfill  \bt{ccccc} If ${\tt P}:[]\to^*{\tt e}:\rho$ &and& ${\tt e}:\rho\Downarrow{\tt e}':\rho'$, &then& 
${\tt e}\Downarrow{\tt e}'$. \\

\hfill  If  ${\tt P}:[]\to^*{\tt e}:\rho$  &and& ${\tt e}:\rho\to{\tt e}':\rho'$, &then& ${\tt e}\to{\tt e}'$.
\et
\vair

\fl We prove both parts of Lemma \ref{abs-super} by course-of-value induction over the size $n=|{\cal D}|$ of a deduction  ${\cal D}$ by Definition \ref{def-evaluate-and-call-relation-environment} of the
assumption 
$${\tt P}:[]\to^*{\tt e}:\rho\land {\tt e}:\rho\Downarrow{\tt e}':\rho'
\mbox{\ or\ }{\tt P}:[]\to^*{\tt e}:\rho\land {\tt e}:\rho\to{\tt e}':\rho'
$$
The deduction size may be 
thought of as the number of 
steps in the computation of 
${\tt e:\rho}\Downarrow{\tt e}':\rho'$ or ${\tt e}:\rho\to{\tt e}':\rho'$ starting from ${\tt P}:[]$.

The induction hypothesis $IH(n)$ is that the Lemma holds for all 
deductions of size not exceeding $n$.
This implies that the Lemma holds for all
calls and evaluations performed in the computation before the last 
conclusion giving 
(${\tt P}:[]\to^*{\tt e}:\rho$ and ${\tt e}:\rho \Downarrow {\tt e}':\rho'$) or (${\tt P}:[]\to^*{\tt e}:\rho$ and ${\tt e}:\rho \to {\tt e}':\rho'$), i.e.,  the Lemma holds for premises of the  rule last applied, and 
{\em for any call and 
evaluation in the computation until then}.

Proof  is by cases on which rule is applied to conclude 
${\tt e:\rho}\Downarrow{\tt e}':\rho'$ or ${\tt e}:\rho\to{\tt e}':\rho'$. 
In all cases we show that some corresponding abstract interpretation rules  
can be applied to 
give the desired conclusion. 

Base cases: Rule (Value), (Operator) and 
(Operand) in the exact semantics (def. \ref{def-evaluate-and-call-relation-environment}) are modeled by axioms (ValueA), 
(OperatorA) and 
(OperandA) in the abstract semantics (def. \ref{def-eval-approx}). These are the same as their 
exact-evaluation 
counterparts, after removal of  environments for  (ValueA) and 
(OperatorA), 
and a premise as well for (OperandA). 
Hence the Lemma holds if one of these rules was the last one applied.

The (Var) rule is, however, rather different from the (VarA) rule.
If  (Var)  was applied to a variable {\tt x} 
then the assumption is (${\tt P}:[]\to^*{\tt x}:\rho$ and 
${\tt x}:\rho\Downarrow {\tt e}':\rho'$).
In this case $x \in dom(\rho)$ 
and
$ {\tt e}':\rho'= \rho(x)$.   
The total size of the deduction 
(of both parts together) 
is $n$. 

Now ${\tt P}:[]\to^*{\tt x}:\rho$ begins from 
the empty environment, and we know all calls are from state to state.
The only possible way {\tt x} can have been bound is 
by a previous use of the (Call) rule, 
the only rule that extends an environment.\footnote{This must have occurred in the part
${\tt P}:[]\to^*{\tt x}:\rho$.}

The premises of the (Call) rule require that 
operator and operand in an application  have previously been evaluated. So it must be the case that there exist ${\tt e}_1{\tt@e}_2:\rho''$ and $\lambda{\tt x.e}_0:\rho_0$ such that 
($P:[]\to^*{\tt e}_1{\tt@e}_2:\rho''$ and 
${\tt e}_1:\rho''\Downarrow \lambda {\tt x.e}_0:\rho_0$ and 
${\tt e}_2 :\rho''\Downarrow {\tt e}':\rho'$) and the size of both deductions are strictly smaller than $n$. 
By the Subexpression Lemma, ${\tt e}_1{\tt@ e}_2 \in subexp({\tt P})$. 
By induction, 
Lemma \ref{abs-super} holds for both ${\tt e}_1:\rho''\Downarrow \lambda {\tt x.e}_0:\rho_0$ and 
${\tt e}_2 :\rho''\Downarrow {\tt e}':\rho'$, so ${\tt e}_1\Downarrow \lambda {\tt x.e}_0$ and 
${\tt e}_2 \Downarrow {\tt e}'$ in the abstract semantics. 
Now we have all premises of rule (VarA),
so we can conclude that ${\tt x} \Downarrow {\tt e}'$ as required.

For remaining rules (Apply) and (Call), when we assume that the Lemma holds for the premises in 
the rule applied to conclude ${\tt e}\Downarrow{\tt e}'$ or ${\tt e}\to{\tt e}'$, 
then this gives us the premises for the corresponding rule for abstract interpretation. 
From this we can conclude the desired result.
\eprf

\section{Proof of Lemma \ref{lem-well-founded}}

\bprf
Define the length $L({\tt e})$ of an expression e by:
$$L({\tt x}) = 1\hair\hair L(\lambda {\tt x}.{\tt e}) = 1 + L({\tt e})\hair\hair 
L({\tt e}_1@{\tt e}_2) = 1 + L({\tt e}_1) + L({\tt e}_2)$$
For any expression {\tt e}, $L({\tt e})$ is a natural number $>0$.  For a program, the length of the initial expression bounds all lengths of occurring expressions.\\

\fl
Define for a state $s$ the height $H(s)$ of the state to be the height of the environment:
$$H({\tt e}:\rho) = max\{(1+H(\rho({\tt x})))\mid {\tt x}\in {\it fv}({\tt e})\}$$
So, $H({\tt e}:[])=0$ the maximum of the empty set, and for any state ${\tt e}:\rho, H({\tt e}:\rho)$ is a natural number $\geq 0$.
Let $>_{lex}$ stand for lexicographic order relation on pairs of natural numbers,
hence $>_{lex}$ is well-founded.
We prove that the relation $\succ$ on states is well-founded
by proving that
${\tt e}_1:\rho_1 \succ {\tt e}_2:\rho_2$
implies
that
$$(H({\tt e}_1:\rho_1),L({\tt e}_1)) >_{lex} (H({\tt e}_2:\rho_2),L({\tt e}_2))$$

First, consider $\succ_1$. Clearly, if ${\tt e}_1:\rho_1 \succ_1 {\tt e}_2:\rho_2$  then
$H({\tt e}_1:\rho_1) > H({\tt e}_2:\rho_2)$. 
Hence even though $L({\tt e}_2)$ might be  larger than $L({\tt e}_1)$, 
it holds that in the lexicographic order  
$(H({\tt e}_1:\rho_1),L({\tt e}_1)) >_{lex} (H({\tt e}_2:\rho_2),L({\tt e}_2))$.

Now, consider $\succ_2$. If ${\tt e}_1:\rho_1 \succ_2 {\tt e}_2:\rho_2$ then
$H({\tt e}_1:\rho_1) \geq H({\tt e}_2:\rho_2)$
and $L({\tt e}_1) > L({\tt e}_2)$, hence
in the lexicographic order
$(H({\tt e}_1:\rho_1),L({\tt e}_1)) >_{lex} (H({\tt e}_2:\rho_2),L({\tt e}_2))$.
Trivially, ${\tt e}_1:\rho_1 = {\tt e}_2:\rho_2$ implies
$(H({\tt e}_1:\rho_1),L({\tt e}_1)) =_{lex} (H({\tt e}_2:\rho_2),L({\tt e}_2))$.

Recall, by definition $\succeq$ is the transitive closure 
of $\succ_1 \cup \succ_2 \cup =$, and $s_1\succ s_2$ 
holds when $s_1\succeq s_2$ and $s_1\neq s_2$.
So, from the derivations above we can conclude that
${\tt e}_1:\rho_1 \succ {\tt e}_2:\rho_2$ implies
$(H({\tt e}_1:\rho_1),L({\tt e}_1)) >_{lex} (H({\tt e}_2:\rho_2),L({\tt e}_2))$,
hence the relation $\succ$ on states is well-founded.

\eprf

\section{Proof of Theorem \ref{graphs_safe}}

\bprf 
For the 
``safety'' theorem we use induction on proofs of 
$s\Downarrow s', G$ or
$s\to s', G$. 
Safety of the constructed graphs  for 
rules (ValueG), (OperatorG) and (OperandG)
is immediate by Definitions \ref{def-safe-call-and-evaluation}  
and \ref{def-value-decrease}. 
\vair

In the following ${\tt x, y, z}$ are variables and $p,q$ can be variables or $\epsilon$.

\vair
The variable lookup rule {\bf (VarG)} yields
${\tt x}:\rho\Downarrow \rho({\tt x}),G$ with
$G= \{ {\tt x}\stackrel{\downarrow}{\to}{\tt y}
         \ |\ {\tt y} \in {\it fv}({\tt e}')\}\cup
\{{\tt x}\stackrel{=}{\to}\epsilon\}$ and
$\rho({\tt x}) = {\tt e}':\rho'$. 
By Definition \ref{def-graph-valuation},
$\overline{{\tt x}:\rho}({\tt x}) 
  = \overline{\rho({\tt x})}(\epsilon)$,
  so arc ${\tt x}\stackrel{=}{\to}\epsilon$   satisfies 
Definition \ref{def-safe-call-and-evaluation}. 
Further, if 
${\tt x}\stackrel{\downarrow}{\to}{\tt y}\in G$ then
${\tt y} \in {\it fv}({\tt e}')$. 
Thus 
$\overline{x:\rho}(x) =\rho({\tt x})= e':\rho'\succ\rho'({\tt y})=\overline{\rho(x)}(y)$ as required.

\vair
The rule  {\bf (CallG)} concludes $s\tosub{c}s', G$, where
$s = {\tt e}_1{\tt@e}_2:\rho$ and 
$s' = {\tt e}_0:\rho_0[{\tt x}\mapsto v_2]$ and
$G=G_1^{-\epsilon/\lambda x.e_0}\cup_{e_0} G^{\epsilon\mapsto{\tt x}}_2$. Its
premises are
${\tt e}_1:\rho\Downarrow\lambda{\tt x.e}_0:\rho_0,G_1$ and
${\tt e}_2:\rho\Downarrow v_2,G_2$. 
We assume inductively that  $G_1$ is safe for
$({\tt e}_1:\rho,\lambda{\tt x.e}_0:\rho_0)$ and that
$G_2$ is safe for $({\tt e}_2:\rho, v_2)$. Let $v_2 = {\tt e}':\rho'$.

We wish to show safety: that 
$p\stackrel{=}{\to}p'\in G$ implies 
$\overline{s}(p) = \overline{s'}(p')$, and  
$p\stackrel{\downarrow}{\to}p'\in G$ implies 
$\overline{s}(p) \succ \overline{s'}(p')$.
By definition of $G_1^{-\epsilon/\lambda x.e_0}$ and $G^{\epsilon\mapsto{\tt x}}_2$,  
$p\stackrel{r}{\to}p'\in 
   G=G_1^{-\epsilon/\lambda x.e_0}\cup_{e_0} G^{\epsilon\mapsto{\tt x}}_2$
breaks into 7 cases:

\vair
\fl {\em Case 1:} 
   ${\tt y}\stackrel{\downarrow}{\to}{\tt z}\in G_1^{-\epsilon/\lambda x.e_0}$
because ${\tt y}\stackrel{\downarrow}{\to}{\tt z}\in G_1$. 
By safety of $G_1$,
$\overline{{\tt e}_1:\rho}({\tt y})\succ\overline{\lambda {\tt x.e}_0:\rho_0}({\tt z})$.
Thus, as required,
$$ \overline{s}({\tt y})
= \overline{{\tt e}_1{\tt@e}_2:\rho}({\tt y}) 
= \overline{{\tt e}_1:\rho}({\tt y})
\succ \overline{\lambda {\tt x.e}_0:\rho_0}({\tt z})
= \overline{{\tt e}_0:\rho_0[{\tt x}\mapsto v_2]}({\tt z})
= \overline{s'}({\tt z})
$$

\vair
\fl {\em Case 2:} 
   ${\tt y}\stackrel{=}{\to}{\tt z}\in G_1^{-\epsilon/\lambda x.e_0}$
because ${\tt y}\stackrel{=}{\to}{\tt z}\in G_1$.
Like Case 1.

\vair
\fl   {\em Case 3:} 
    ${\tt y}\stackrel{\downarrow}{\to}\epsilon\in G_1^{-\epsilon/\lambda x.e_0}$
because ${\tt y}\stackrel{r}{\to}{\epsilon}\in G_1$, 
then $x\notin {\it fv}({\tt e}_0)$ by the definition of $G_1^{-\epsilon/\lambda x.e_0}$ and then ${\tt e}_0:\rho_0[{\tt x}\mapsto v_2]={\tt e}_0:\rho_0$.
By safety of $G_1$,
$\overline{{\tt e}_1:\rho}({\tt y})\succeq\overline{\lambda {\tt x.e}_0:\rho_0}({\epsilon})= \lambda {\tt x.e}_0:\rho_0$.
Thus, as required,
$$ \overline{s}({\tt y})
= \overline{{\tt e}_1{\tt@e}_2:\rho}({\tt y}) 
= \overline{{\tt e}_1:\rho}({\tt y})
\succeq \lambda {\tt x.e}_0:\rho_0
\succ {\tt e}_0:\rho_0 
= \overline{s'}({\epsilon})
$$

\vair
\fl {\em Case 4:} 
   $\epsilon\stackrel{\downarrow}{\to}{p}\in G_1^{-\epsilon/\lambda x.e_0}$
because $\epsilon\stackrel{r}{\to}{p}\in G_1$. Then it holds that either ${p}$ is a variable-name or  ${\tt x}\notin {\it fv}({\tt e}_0)$. 
Now $\epsilon$ in $G_1$ refers to ${\tt e}_1:\rho$, so
${\tt e}_1:\rho \succeq\overline{\lambda {\tt x.e}_0:\rho_0}({ p})$  by safety of $G_1$.
Thus, as required,
$$ \overline{s}(\epsilon)
= {\tt e}_1{\tt@e}_2:\rho
\succ {\tt e}_1:\rho
\succeq\overline{\lambda {\tt x.e}_0:\rho_0}({p})
\succeq \overline{{\tt e}_0:\rho_0[{\tt x}\mapsto v_2]}({ p})
= \overline{s'}({p})
$$

\vair
\fl {\em Case 5:} 
${\tt y}\stackrel{\downarrow}{\to}{\tt x}\in G$ because ${\tt x}\in {\it fv}({\tt e}_0)$ and 
${\tt y}\stackrel{\downarrow}{\to}{\tt x}\in G^{\epsilon\mapsto{\tt x}}_2$
because ${\tt y}\stackrel{\downarrow}{\to}\epsilon\in G_2$. 
By safety of $G_2$, 
$\overline{{\tt e}_2:\rho}({\tt y})\succ \overline{v_2}(\epsilon)$.
Thus, as required,
$$ \overline{s}({\tt y})=\overline{{\tt e}_1{\tt@e}_2:\rho}({\tt y}) 
= \overline{e_2:\rho}({\tt y})
\succ \overline{v_2}(\epsilon)
= \overline{{\tt e}_0:\rho_0[{\tt x}\mapsto v_2]}({\tt x})
= \overline{s'}({\tt x})
$$

\vair
\fl {\em Case 6:} 
${\tt y}\stackrel{=}{\to}{\tt x}\in G$ because $x\in {\it fv}(e_0)$ and
${\tt y}\stackrel{=}{\to}{\tt x}\in G^{\epsilon\mapsto{\tt x}}_2$
because ${\tt y}\stackrel{=}{\to}{\epsilon}\in G_2$. 
Like Case 5.

\vair 
\fl {\em Case 7:}
$\epsilon\stackrel{\downarrow}{\to}{\tt x}\in G$ because ${\tt x}\in {\it fv}({\tt e}_0)$ and
$\epsilon\stackrel{\downarrow}{\to}{\tt x}\in G^{\epsilon\mapsto{\tt x}}_2$
because $\epsilon\stackrel{r}{\to}\epsilon\in G_2$.
By safety of $G_2$, 
$\overline{{\tt e}_2:\rho}(\epsilon)= {\tt e}_2:\rho$.
Thus, as required,
$$ \overline{s}(\epsilon)
= {\tt e}_1{\tt@e}_2:\rho
\succ {\tt e}_2:\rho
\succeq \overline{v_2}(\epsilon)
= \overline{\rho_0[{\tt x}\mapsto v_2]}({\tt x})
= \overline{s'}({\tt x})
$$

\vair
The rule {\bf (ApplyG)} concludes  $s\Downarrow  v, G';G$ 
from premises
$s\tosub{c}s', G'$
and
$s'\Downarrow v,G$, where
$s = {\tt e}_1{\tt@e}_2:\rho$ and $s' = {\tt e}':\rho'$.
We assume inductively 
that $G'$ is safe for $(s,s')$
and $G$ is safe for $(s',v)$.
Let $G_0 = G';G$.

We wish to show that $G_0$ is safe: that 
$p\stackrel{=}{\to}q\in G_0$ implies
$\overline{s}(p) = \overline{v}(q)$, and
$p\stackrel{\downarrow}{\to}q\in G_0$ implies
$\overline{s}(p)\succ\overline{v}(q)$ ($p, q$ can be variables or $\epsilon$).
First, consider the case $p\stackrel{=}{\to}q\in G_0$.
Definition \ref{def-size-change-terminology} 
implies
$p\stackrel{=}{\to}p'\in G'$
and $p'\stackrel{=}{\to}q\in G$
for some $p'$. Thus by the inductive assumptions we have
$\overline{s}(p) = \overline{s'}(p') = \overline{v}(q)$, as required.

Second, consider the case 
$p\stackrel{\downarrow}{\to}q\in G_0$.
Definition \ref{def-size-change-terminology} 
implies
$p\stackrel{r_1}{\to}p'\in G'$
and $p'\stackrel{r_2}{\to}q\in G$
for some $p'$, where either one or both of $r_1,r_2$ are $\downarrow$. 
By the inductive assumptions we have
$\overline{s}(p) \succeq \overline{s'}(p')$ and 
$\overline{s'}(p') \succeq \overline{v}(q)$, and 
one or both of
$\overline{s}(p) \succ \overline{s'}(p')$ and 
$\overline{s'}(p') \succ \overline{v}(q)$ hold.
By Definition of $\succ$ and $\succeq$ this implies that 
$\overline{s}(p) \succ  \overline{v}(q)$, as required.

\eprf

\section{Proof of Lemma \ref{abs-graphs}}

\bprf 
The rules are the same as in Section \ref{def-eval-approx}, 
only extended with size-change graphs. 
We need to add to Lemma \ref{abs-super} that the size-change graphs 
generated for
calls and evaluations can also be generated by the abstract interpretation. 
The proof is by cases on which rule is applied to conclude 
${\tt e}\Downarrow{\tt e}',G$ or ${\tt e}:\rho\to{\tt e}':\rho',G$. 

We build on Lemma \ref{abs-super}, and we saw in the proof of this that 
in abstract interpretation we can always use a rule corresponding to 
the one used in exact computation to prove corresponding steps. 
The induction hypothesis is that the Lemma holds for the premises of the rule 
in exact semantics. 

Base case (VarAG): By Lemma \ref{abs-super} 
we have ${\tt x}:\rho \Downarrow {\tt e}':\rho'$ implies 
${\tt x} \Downarrow {\tt e}'$. The size-change graph built in (VarAG) 
is derived in the same way from {\tt x} and ${\tt e}'$ 
as in rule (VarG), and they will therefore be identical.

For other call- and evaluation rules without premises, 
the abstract evaluation rule 
is as the exact-evaluation rule, only with environments removed, and 
the generated size-change graphs 
are not influenced by environments. Hence the Lemma will hold if these rules are applied.

For all other rules in a computation: When we know that 
Lemma \ref{abs-super} holds and 
assume that Lemma \ref{abs-graphs} hold for the premises, then we 
can conclude that 
if this rule is applied, then Lemma \ref{abs-graphs} holds by the 
corresponding rule from 
abstract interpretation. 
\eprf
\section{Proof of Lemma \ref{simulation}}
\bprf By induction on the tree for the proof of evaluation or call in the pure $\lambda$-calculus.\\
Possible cases of the structure of ${\tt e}':\rho'$ and ${\tt e}:\rho$ in $S$-related states:\smallskip

\bt{rlccrlccrl}
$({\tt x}:\rho',$&${\tt x}:\rho)$& & &
$(\lambda {\tt x.e}':\rho',$&$\lambda {\tt x.e}:\rho)$& & &
$({\tt e}_1'{\tt @e}_2':\rho',$&${\tt e}_1{\tt @e}_2:\rho)$\\
$({\tt A}:\rho',$&${\tt x}:\rho)$& & &
$({\tt A}:\rho',$&$\lambda {\tt x.e}:\rho)$& & &
$({\tt A}:\rho',$&${\tt e}_1{\tt @e}_2:\rho)$\\
\et
\medskip

Base cases, evaluations and calls in pure $\lambda$-calculus by rules without premisses.
\medskip

Case $S({\tt x}:\rho',{\tt x}:\rho)$: No calls from ${\tt x}:\rho$.\\ 
\fl (Var)-rule, ${\tt x}:\rho \Downarrow \rho({\tt x})= {\tt e}_0:\rho_0,\ \{ {\tt x}\stackrel{=}{\to}{\epsilon}\}
  \cup\{ {\tt x}\stackrel{\downarrow}{\to}{\tt y}
         \ |\ {\tt y} \in {\mathit fv}({\tt e}_0)\}$ and ${\tt x}:\rho' \Downarrow \rho'(x)={\tt e}_0':\rho_0',\ \{ {\tt x}\stackrel{=}{\to}{\epsilon}\}
  \cup\{ {\tt x}\stackrel{\downarrow}{\to}{\tt y}
         \ |\ {\tt y} \in {\mathit fv}({\tt e}_0')\}$. Beginning from $S$-related states, by defintion of the relation $S$ we have $S(\rho'({\tt x}),\rho({\tt x}))$ and $\mathit{fv}({\tt e}_0')\supseteq \mathit{fv}({\tt e}_0)$. $source(G')= source(G)$ and the generation of size-change graphs gives that the restriction of $G'$ to $target(G)$ equals $G$, hence $T(G',G)$. \\
\\
Case $S(\lambda {\tt x.e}':\rho',\lambda {\tt x.e}:\rho)$: No calls from $\lambda {\tt x.e}:\rho$. \\
(Value)-rule, 
$\lambda {\tt x.e}:\rho\Downarrow \lambda {\tt x.e}:\rho,{\it id}^=_{\lambda{\tt x.e}}$ 
and $\lambda {\tt x.e}':\rho'\Downarrow \lambda {\tt x.e}':\rho',{\it id}^=_{\lambda{\tt x.e'}}$. $T({\it id}^=_{\lambda{\tt x.e'}},{\it id}^=_{\lambda{\tt x.e}})$.\\
\\
Case $S({\tt e}_1'{\tt @e}_2':\rho',{\tt e}_1{\tt @e}_2:\rho)$:\\ 
(Operator)-rule, ${\tt e}_1{\tt @e}_2:\rho\tosub{r} {\tt e}_1:\rho,{\it id}^\downarrow_{e_1}$ and 
${\tt e}_1'{\tt @e}_2':\rho'\tosub{r} {\tt e}_1':\rho',{\it id}^\downarrow_{e_1'}$.Beginning from $S$-related states, by defintion of the relation $S$ we have $S({\tt e}_1':\rho',{\tt e}_1:\rho)$. Then $T({\it id}^\downarrow_{e_1'},{\it id}^\downarrow_{e_1})$\\
\\
Case $S({\tt A}:\rho',{\tt x}:\rho)$: \\
(Var)-rule: ${\tt x}:\rho\Downarrow \rho({\tt x})={\tt e}_0:\rho_0,G$ where $G=\{ {\tt x}\stackrel{=}{\to}{\epsilon}\}
  \cup\{ {\tt x}\stackrel{\downarrow}{\to}{\tt y}
         \ |\ {\tt y} \in {\mathit fv}({\tt e}_0)\}$.  
By the definition of $S$ we must have ${\tt A}\Rightarrow_\Gamma^*x$. This againg by lemma \ref{prod-to-form} gives that we must have ${\tt A}::={\tt x}$. Then ${\tt A}:\rho'\tosub{n}{\tt x}:\rho',\ id^=_x$ by (Gram)-rule, and we have $S({\tt x}:\rho',{\tt x}:\rho)$. Also ${\tt x}:\rho'\Downarrow \rho'({\tt x}) ={\tt e}_0':\rho_0',G''$ where $G''=\{ {\tt x}\stackrel{=}{\to}{\epsilon}\}
  \cup\{ {\tt x}\stackrel{\downarrow}{\to}{\tt y}
         \ |\ {\tt y} \in {\mathit fv}({\tt e}_0')\}$ by (Var)-rule. 
The edges in $G''$ are the same as the edges in $G'=id^=_x;G''$.
Hence by (Result)-rule $A\Downarrow \rho'({\tt x}) ,G'$. 
As before $S(\rho'({\tt x}),\rho({\tt x}))$ and $T(G',G)$.\\
\\
Cases $S({\tt A}:\rho',\lambda {\tt x.e}:\rho)$ with (Value)-rule, and $S({\tt A}:\rho',{\tt e}_1{\tt @e}_2:\rho)$ with (Operator)-rule: Similarly by use of lemma \ref{prod-to-form} and reasoning as above. We will use the rules (Gram)(Value) (Result) and (Gram)(Operator) respectively, where (Value) and (Operator) do not have premises.\\
\\
Step cases.\\
\\
Case $S({\tt e}_1'{\tt @e}_2':\rho',{\tt e}_1{\tt @e}_2:\rho)$. ${\tt e}_1{\tt @e}_2:\rho\tosub{d} {\tt e}_2:\rho,{\it id}^\downarrow_{{\tt e}_2}$ by (Operand)-rule.
It follows from the definition of $S$ that also $S({\tt e}_1':\rho',{\tt e}_1:\rho)$ hence by IH since ${\tt e}_1:\rho\Downarrow$ then also ${\tt e}_1':\rho'\Downarrow$ and then ${\tt e}_1'{\tt @e}_2':\rho'\tosub{d} {\tt e}_2':\rho',{\it id}^\downarrow_{{\tt e}_2'}$ and by the definition of $S$ we have $S({\tt e}_2':\rho',{\tt e}_2:\rho)$, $T({\it id}^\downarrow_{{\tt e}_2'},{\it id}^\downarrow_{{\tt e}_2})$.\\
\\
The next case is the one that requires the most consideration to see that we stay within the $T$-relation. Assume we know for graphs $\tilde G',\tilde G$, that the restriction of $\tilde G'$ to source and target of $\tilde G$ is a subset of $\tilde G$. Notice, if $x,y\in source(\tilde G')\setminus source(\tilde G)$ and  ${\tt x},{\tt z} \in target(\tilde G')\setminus target(\tilde G)$, then for testing $T(\tilde G',\tilde G)$ we only need to look at which edges leaves from ${\tt x,y}$, we do not need to care about if other edges goes into ${\tt x,z}$.\\
\\
Case $S({\tt e}_1'{\tt @e}_2':\rho',{\tt e}_1{\tt @e}_2:\rho)$. ${\tt e}_1{\tt @e}_2:\rho\tosub{c} {\tt e}_0:\rho_0[{\tt x}\mapsto v_2], G_1^{-\epsilon/\lambda {\tt x.e}_0}\cup_{{\tt e}_0} G^{\epsilon\mapsto{\tt x}}_2$ by (Call)-rule, where we have the premises ${\tt e}_1:\rho\Downarrow\lambda{\tt x.e}_0:\rho_0,G_1$ and ${\tt e}_2:\rho\Downarrow v_2,G_2$.

It follows from the definition of $S$ that also $S({\tt e}_1':\rho',{\tt e}_1:\rho)$ and $S({\tt e}_2':\rho',{\tt e}_2:\rho)$. Hence by IH since ${\tt e}_1:\rho\Downarrow \lambda {\tt x.e}_0:\rho_0,G_1$ then also ${\tt e}_1':\rho'\Downarrow v,G_1'$ where $T(G_1',G_1)$ and $S(v,\lambda {\tt x.e}_0:\rho_0)$. Then by definition of values, relations $\Rightarrow_\Gamma^*$ and $S$ we must have $v = \lambda {\tt x.e}_0':\rho_0'$. Also by IH since ${\tt e}_2:\rho\Downarrow v_2,G_2$ then also ${\tt e}_2':\rho'\Downarrow v_2',G_2'$ where $T(G_2',G_2)$ and $S(v_2',v_2)$. Then we have the premises to conclude ${\tt e}_1'{\tt @e}_2':\rho'\tosub{c} {\tt e}_0':\rho_0'[{\tt x}\mapsto v_2'], G_1'^{-\epsilon/\lambda x.e_0'}\cup_{e_0'} G_2'^{\epsilon\mapsto{\tt x}}$. By definition of $S$ we have $S({\tt e}_0':\rho_0'[{\tt x}\mapsto v_2'], {\tt e}_0:\rho_0[{\tt x}\mapsto v_2])$. 
We notice that ${\tt x}\notin \mathit{fv}(\lambda {\tt x.e}_0')$ and therefore $(p\stackrel{r}{\to}{\tt x})\notin G_1'$.\\
We consider different possibilities for the generated graphs:

If ${\tt x}\in \mathit{fv}({\tt e}_0')$ but ${\tt x}\notin \mathit{fv}({\tt e}_0)$ then we can have some extra edges going to ${\tt x}$ in extended semantics where we will have no edges to ${\tt x}$ in pure semantics because ${\tt x}$ is not in the target, but this is acceptable in the $T$-relation. There can also be some extra edges going to $\epsilon$ in pure semantics where no edges go to $\epsilon$ in exact semantics, but as $\epsilon$ is within the codomain in pure semantics, this is also acceptable in the $T$-relation. Since $T(G_1',G_1)$ it will still hold that $T(G_1'^{-\epsilon/\lambda x.e_0'}\cup_{e_0'} G_2'^{\epsilon\mapsto{\tt x}}, G_1^{-\epsilon/\lambda x.e_0}\cup_{e_0} G_2^{\epsilon\mapsto{\tt x}})$. 

If ${\tt x}\in \mathit{fv}({\tt e}_0)$ then also ${\tt x}\in \mathit{fv}({\tt e}_0')$ and if ${\tt x}\notin \mathit{fv}({\tt e}_0')$ then ${\tt x}\notin \mathit{fv}({\tt e}_0)$, in these cases since $T(G_1',G_1)$ and $T(G_2',G_2)$ also $T(G_1'^{-\epsilon/\lambda x.e_0'}\cup_{e_0'} G_2'^{\epsilon\mapsto{\tt x}}, G_1^{-\epsilon/\lambda x.e_0}\cup_{e_0} G_2^{\epsilon\mapsto{\tt x}})$. \\
\\
Case $S({\tt A}:\rho',{\tt e}_1{\tt @e}_2:\rho)$ with ${\tt e}_1{\tt @e}_2:\rho\tosub{d} {\tt e}_2:\rho,{\it id}^\downarrow_{{\tt e}_2} $ by(Operand)-rule. By the definition of $S$ we must have ${\tt A}\Rightarrow_\Gamma^*{\tt e}_1{\tt @e}_2$. This againg by lemma \ref{prod-to-form} gives that we must have ${\tt A}::= {\tt e}_1'{\tt @e}_2'$. Then ${\tt A}:\rho'\tosub{n}{\tt e}_1'{\tt @e}_2':\rho',\ id^=_{ e_1'@e_2'}$ by (Gram)-rule, and we have $S({\tt e}_1'{\tt @e}_2':\rho', {\tt e}_1{\tt @e}_2:\rho)$. Then we have seen that ${\tt e}_1'{\tt @e}_2':\rho'\tosub{d} {\tt e}_2':\rho',{\it id}^\downarrow_{{\tt e}_2'}$ with $S({\tt e}_2':\rho',{\tt e}_2:\rho)$, $T({\it id}^\downarrow_{{\tt e}_2'},{\it id}^\downarrow_{{\tt e}_2})$ and we have that the edges of ${\it id}^\downarrow_{{\tt e}_2'}$ are the same as the edges of ($id^=_{ e_1'@e_2'};{\it id}^\downarrow_{{\tt e}_2'}$) hence $T((id^=_{ e_1'@e_2'};{\it id}^\downarrow_{{\tt e}_2'}),{\it id}^\downarrow_{{\tt e}_2} )$.\\
\\
Case $S(A:\rho',e_1@e_2:\rho)$ with (Call)-rule $e_1@e_2:\rho\tosub{c} e_0:\rho_0[x\mapsto v_2], G$: Similarly as before we have $A:\rho'\tosub{n}e_1'@e_2':\rho',\ id^=_{ e_1'@e_2'}$ by (Gram)-rule, and we have $S(e_1@e_2:\rho, e_1'@e_2':\rho')$. We can now use the derivation above and with the notation from above we have $ e_1'@e_2':\rho'\tosub{c} e_0':\rho_0'[x\mapsto v_2'], G'$ with $S(e_0':\rho_0'[x\mapsto v_2'], e_0:\rho_0[x\mapsto v_2])$ and $T(G',G)$. Looking into the derivation of $G'$ we find that the edges of $G'$ are the same as the edges of ($id^=_{ e_1'@e_2'};G'$).\\
\\
Case $S({\tt e}':\rho',{\tt e}:\rho)$, ${\tt e}:\rho\Downarrow v,G$ by (Result)-rule, where we have the premises ${\tt e}:\rho\tosub{c} {\tt e}_s:\rho_s,G_s$ and ${\tt e}_s:\rho_s\Downarrow v,G_v$, $G=G_s;G_v$: 
By IH since $e:\rho\tosub{c} {\tt e}_s:\rho_s,G_s$ then ${\tt e}':\rho'\tosub{n}^j s:\rho'\tosub{c} {\tt e}_s':\rho_s',G_s'$ with $S({\tt e}_s':\rho_s',{\tt e}_s:\rho_s)$, and $T(G_s',G_s)$, $j\in\{ 0,1\}$. Again by IH since ${\tt e}_s:\rho_s\Downarrow v,G_v$ then ${\tt e}_s':\rho_s'\Downarrow v',G_v'$ with $S(v,v')$ and $T(G_v',G_v)$. Let $G'=G_s';G_v'$ then $T(G',G)$. If $j=0$ we have the premises to conclude ${\tt e}':\rho'\Downarrow v',G'$. If $j=1$ by lemma \ref{push-S} we have $S(s:\rho',{\tt e}:\rho)$ and we have the premises to conclude $s:\rho'\Downarrow v',G$, and by applications of (Result)-rule once more in the extended semantics we can also conclude ${\tt e}':\rho'\Downarrow v', id_s^=;G'$ where the edge set of $G'$ is the same as the edge set of $id_s^=;G'$.
\eprf

\nocite{*}
\bibliographystyle{alpha}

\end{document}